\documentclass{aa}
\usepackage{graphicx}
\usepackage{txfonts}
\usepackage{natbib}
\usepackage[figuresright]{rotating}
\usepackage{aalongtable,lscape}

\def\cyg{{Cygnus\,OB2\,\,}}
\def\chandra{{\em Chandra\,\,}}

\newcommand{\Av}{\mbox{A$_{\rm v}$\,}}
\begin{document}
   \title{Unveiling the Cygnus\,OB2 stellar population with Chandra}
   
   \author{J.F. Albacete Colombo\inst{1}
     \and
     E. Flaccomio\inst{1}
     \and
     G. Micela\inst{1}
     \and
     S. Sciortino\inst{1}
     \and
     F. Damiani\inst{1}	
   }
   
   \offprints{J.F.A.C. -  
email: facundo@astropa.unipa.it}
   
   \institute{INAF - Osservatorio Astronomico di Palermo,
     Piazza del Parlamento 1, I-90134, Palermo, Italy.\\
     \email{facundo@astropa.unipa.it}
   }
   
   \date{Received -----; accepted -----}
   
   
   \abstract
{}
{The aim of this work is to identify the so far unknown low mass stellar 
population of the $\sim2$Myr old \cyg star forming region, and to investigate 
the X-ray and near-IR stellar properties of its members.}
{We analyzed a 97.7 ksec \chandra ACIS-I observation pointed at the core of 
the \cyg region. Sources were detected using the {{\sc PWDetect}} code and 
were then positionally correlated with optical and near-IR catalogs from the 
literature. Source events were then extracted with the {{\sc Acis Extract}} 
package.  X-ray variability was characterized through the Kolmogorov-Smirnov
test and  spectra were fitted using absorbed thermal plasma models.}
{We detected 1003 X-ray sources. Of these, 775 have near-IR counterparts and
are expected to be almost all associated  with \cyg members. From near-IR
color-color  and color-magnitude diagrams we estimate a typical absorption
toward \cyg of  \Av\,$\approx$\,7.0\,mag. Although the region is young, very
few stars ($\sim$4.4\%)  show disk-induced excesses in the near-IR. X-ray
variability is detected in $\sim$13\% of  the sources, but this fraction
increases, up to 50\%, with increasing source  statistics. Flares account for
at least 60\% of the variability. Despite being  generally bright, all but 2 of
the 26 detected O- that early B-type stars are not significantly variable. 
Typical X-ray spectral parameters are log N$_{\rm H}$\,$\sim$\,22.25 
(cm$^{-2}$) and kT\,$\sim$\,1.35 keV with 1$\sigma$ dispersion of 0.2~dex and
0.4~keV,  respectively. Variable and flaring sources have harder spectra with
median kT\,=3.3 and 3.8 keV, respectively.  OB stars are typically softer
(kT\,$\sim$\,0.75 keV). X-ray luminosities range between  $10^{30}$ and
$10^{31}$ erg\,s$^{-1}$ for intermediate- and low-mass stars, and
2.5$\times10^{30}$ and between 6.3$\times10^{33}$ erg\,s$^{-1}$  for OB stars.}
{The \cyg region has a very rich population of low-mass X-ray emitting stars. 
Circumstellar disks seem to be very scarce. X-ray variability is essentially
related  to the magnetic activity of low-mass stars (M/M$_\odot$$\sim$0.5 to
3.0) which display X-ray activity levels comparable to those of 
Orion Nebular Cluster (ONC) sources in the same mass range. }

\keywords{stars: formation -- stars: early-type -- stars: pre-main-sequence --
		Galaxy: globular clusters and OB associations: 
     individual: Cygnus OB2 -- X-rays: stars\\
    		On-line material: machine readable tables, color figures.
}
  
\maketitle

\section{Introduction}

The \cyg association is one of the richest star forming regions in the Galaxy,
containing a large population of O- and B-type stars, some of which are among
the most massive stars known \citep{1991MNRAS.249....1T}. For its richness \cyg
has been in the past considered a ``young globular cluster''
\citep{2000A&A...360..539K}, although uncertainties remain on the real size of
total cluster population \citep{2003ApJ...597..957H}.  At a distance of 1.45
kpc \citep[DM=10.80,][]{2003ApJ...597..957H}, \cyg lies behind the Great Cygnus
Rift and is affected by large and non-uniform extinction, \Av: 5 to 15. Its
distance and absorption have so far hindered optical studies of intermediate-
and  low- mass stars in the region. The position of known massive OB stars in
the HR diagram of \cyg  suggests an age between 1 and 3 Myr, with $\sim$ 2 Myr
as the most probable value \citep{1991AJ....101.1408M,2002A&A...390..945K} .

The \cyg region played a remarkable role in the history of stellar X-ray
astronomy as an early X-ray observation, performed with the {\it Einstein}
satellite and analyzed by \citet{1979ApJ...234L..51H}, first revealed that
early-type stars are intense X-ray emitters.  This work was also the first to
suggest the idea that X-rays in OB stars are produced in shocks within stellar
winds, the starting point for the X-ray emission model of
\citet{1980ApJ...241..300L}. Subsequent {\it ROSAT}, {\it ASCA}, and {\it
Chandra}-HETG observations of massive stars in the \cyg region have improved
our understanding of the mechanisms responsible for the X-ray emission of
O-type stars
\citep{2004ApJ...616..542W,1998ApJS..118..217W,1996PASJ...48..813K}.

These latter studies of the region have focused on the four most known OB-type
stars of the region, Cyg\#5, 8, 9, and 12, but have left the X-ray properties
of the other high-, intermediate-, and low-mass stars in the region unexplored.
Indeed the intermediate/low-mass population has not even been identified yet.
Because of the much higher X-ray luminosity of young stars (e.g. log($L_{\rm
x}$)\,$\sim$\,30\,-\,31 erg/s, for M\,=\,1.0\,-\,3.0\,$M_\odot$) with respect
to older field stars \citep[e.g.][]{2005ApJS..160..353G}, X-ray observations
have proved an efficient  tool to select likely members of young Star Forming
Regions (SFR). Thanks to its high sensitivity and spatial resolution, the
Advanced  CCD Imaging Spectrometer (ACIS) camera on board of the {\it Chandra} 
satellite \citep{2003SPIE.4851...28G} is particularly suited for this  kind of
studies in crowded stellar fields such as \cyg. 

In this paper we present the analysis of a deep ($\sim$\,97.7 ks) {\it Chandra}
ACIS observation of \cyg. In \S 2 we introduce our X-ray observation and
describe our data calibration and reduction procedure. In section 3 we describe
source detection, photon extraction and first characterization of X-ray spectra
via hardness ratios. In \S 4 we characterize our sources by cross-identifying
them with available optical and near-IR catalogs. Section 5 deals with X-ray
variability while in \S 6 the results of the X-ray spectral analysis is
presented. In \S 7 we discuss out results concerning the X-ray characteristics
of high-, intermediate- and low-mass stars of the region. Finally, in \S 8 we
summarize our results and draw our conclusions.

\section{The X-ray observation}

\begin{figure*}[ht] \centering
\includegraphics[width=15cm,angle=0]{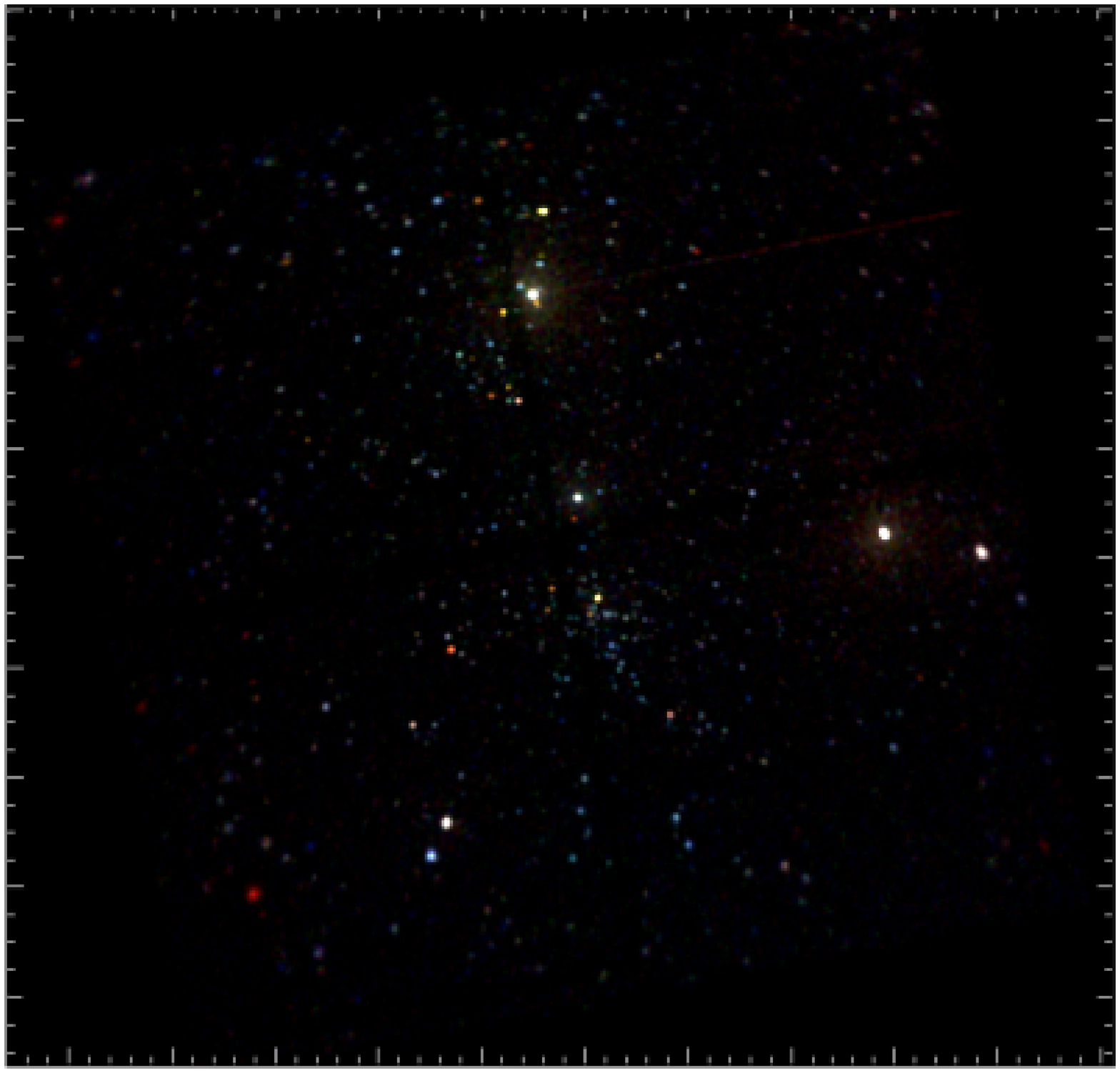}
\includegraphics[width=15cm,angle=0]{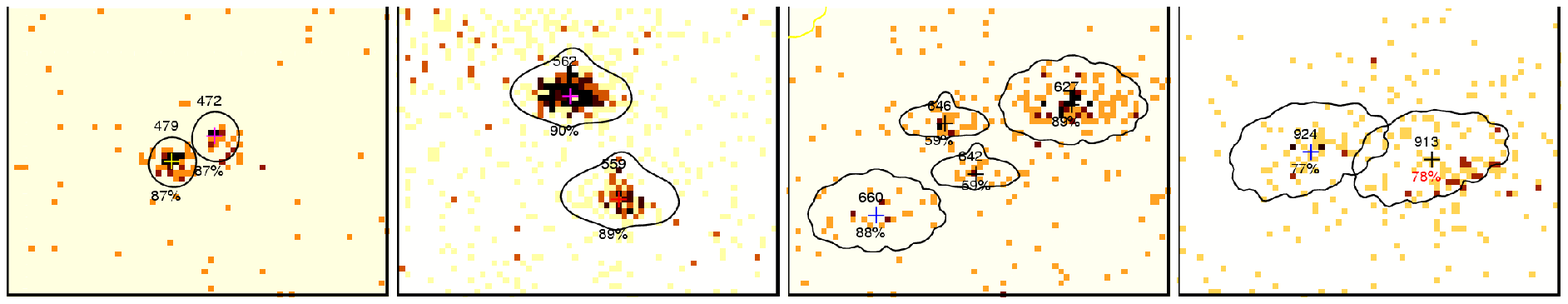} \caption{Upper panel:
color-coded ACIS-I image of the 17'\,$\times$\,17' field in \cyg  (see color
version in the electronic  edition). Kernel smoothing was applied to highlight
point sources. Energy bands for the RGB image are [0.5:1.7], [1.7:2.8], and
[2.8:8.0] keV for the red, green, and blue colors, respectively. The four
panels at the bottom show four  30"\,$\times$\,20" areas of the ACIS field at
four different distances from the optical axis: from left to right: 0.7, 4.7,
5.4, and 7 arcmin. They illustrate the dependence of the PSF quality on
off-axis and show (black continuous lines) the source photon extraction regions
established with {\sc ACIS Extract}. Figures next to these contours indicate
the source identification numbers (from Table 1) and the encircled energy
fractions.} \label{acis} \end{figure*}

\cyg was observed with the ACIS detector on board the  {\it Chandra X-ray
Observatory} (CXO) \citep{2002PASP..114....1W} on 2004 January 16  (Obs.Id.
4511; PI of the observation: E. Flaccomio). The total exposure time was 97.7
ksec.  The data were acquired in {\sc very faint} mode, to ease filtering of
background events, with six CCD turned on, the four comprising the ACIS-I array
[0,1,2,3], plus CCDs 6 and 7, part of ACIS-S. However, these latter two CCDs
will not be used in the following because of the much degraded point spread
function (PSF) and reduced effective area resulting from their large distance
from the optical axis. The ACIS-I 17'$\times$17' field of view (FOV) is covered
by 4 chips each with 1024$\times$1024 pixels (scale 0.49"\,px$^{-1}$). The
observation was pointed toward R.A.=20$^{\rm h}$\,33$^{\rm m}$\,12.2$^{\rm s}$ 
and DEC=+41$^\circ$\,15'\,00.7'',  chosen so to maximize the number of stars in
the FOV and to keep two source-dense regions close to the optical axis, where
the PSF is sharper. Figure \ref{acis} shows \cyg as seen in X-rays by our
ACIS-I observation.

\subsection{Data reduction}

Data reduction, starting with the Level 1 event list provided by the pipeline
processing at the CXO, was performed using {\sc CIAO
3.2.2}\footnote{http://cxc.harvard.edu/ciao/} and the {\sc CALDB 3.1.0} set of
calibration files. We produced a level 2 event file using the {\sc
acis\_process\_event} CIAO task, taking advantage of the VF-mode enhanced
background filtering, and retaining only events with grades\,=\,0,2,3,4,6 and
status=0. Photon energies were corrected for the time dependence of the energy
gain using the {\sc corr\_tgain} CIAO task. Intervals of background flaring
were searched for, but none were found. We will hereafter assume a constant
background.  To improve sensitivity to faint sources, given the spectrum of the
background and that of typical sources, we filtered out events outside the
[500:8000] eV energy band.

An exposure map, needed by the source detection algorithm and to renormalizes
source count-rates, was calculated with the {\sc CIAO} tool {\sc mkexpmap}
assuming a monochromatic spectrum (kT\,=\,2.0 keV),
\footnote{http://asc.harvard.edu/ciao/download/doc/expmap\_intro.ps}.

\section{Analysis}

In this section we describe the first steps taken for the analysis of the
ACIS-I data. We discuss the detection of X-ray point sources (\S
\ref{sect:detection}), the definition of source and background event lists (\S
\ref{sect:extraction}), and a first spectral characterization of source spectra
through hardness ratios (\S \ref{sect:hardratio}).

\subsection{X-ray Source Detection}
\label{sect:detection}

Source detection was performed with the Palermo Wavelet Detection code,
PWDetect\footnote{See http://www.astropa.unipa.it/progetti\_ricerca/PWDetect}
\citep{1997ApJ...483..370D}, on the level 2 event list restricted to the 
[500:8000] eV energy band. PWDetect analyzes the data at different spatial
scales, allowing the detection of both point-like and moderately extended
sources, and efficiently resolving close sources pairs.   The most important
input parameter is the detection threshold (SNR), which we establish from the
relationship between threshold, background level of the observation, and
expected number of spurious detections due to Poisson noise, as determined from
extensive simulations of source-free fields \citep[cf.][]{1997ApJ...483..350D}.
The background level was determined with the {\sc background} command in the
{\sc ximage}\footnote{http://heasarc.nasa.gov/xanadu/ximage/ximage.html}
package. The method amounts to dividing the image into equal-size boxes,
discarding those that, according to several statistical criteria, are
contaminated by sources, and finally computing the mean level of the remaining
ones. We obtain that our ACIS-I observation comprises a total of about
1.58$\times$10$^5$  background photons. This background level translates into a
SNR threshold of 5.2 if we decide to accept one spurious detection in the FOV,
or into SNR\,$>$\,4.5 if we decide to accept 10 spurious detections.  The first
choice results in the detection of 868 sources and the second one in 1054
sources. By accepting an extra $\sim$9 spurious detections in the FOV we thus
gain about 177 new reliable sources. Considering moreover that 10 spurious
sources amounts to only $\sim$1\% of the total number of detections, we decided
to adopt the second less conservative criterion.

After a careful visual inspection of the initial source list, we rejected a
total of 51 detections which we considered spurious:  39 were produced by
different instrumental artifacts (i.e. out-of-time events, CCD gaps,
detector edges, etc) that were not included in the simulations of source-free
fields used to establish the SNR threshold, but that are easily recognized. The
remaining 12 were multiple detections of the same sources with different
spatial scales.\footnote{This effect occurred for only 3 sources at large
off-axis angles, where the PSF is particularly elongated and thus 
significantly different from the symmetric PSF assumed by PWDetect. Note that,
quite obviously, this effect is also not included in the simulations of
source-free fields.} In total, our final list of X-ray sources in the \cyg
region contains 1003 X-ray detections, 99\% of which are expected to be real. 
The first 7 columns of Table \ref{AEphot} list, for each source: a running
source  number, name (according to CXC naming 
convention\footnote{http://cxc.harvard.edu/cdo/scipubs.html}), sky position
(R.A. and Dec.) with relative uncertainty,  off-axis angle ($\theta$),
significance of the detection (Sig.).

\subsection{Photon Extraction}
\label{sect:extraction}

For further analysis of the detected sources, we used {\sc ACIS
Extract}\footnote{http://www.astro.psu.edu/xray/docs/TARA/ae\_users\_guide.html}
(AE) v3.79 \citep{acisextract}, an IDL based package that makes use of
TARA\footnote{http://www.astro.psu.edu/xray/docs/TARA/}, CIAO and
FTOOLS\footnote{http://heasarc.gsfc.nasa.gov/docs/software/ftools/}.

The optimal extraction of point source photons is complicated by the
non-Gaussian shape of the PSF and by its strong non-uniformity across the ACIS
field of view. In particular the width and asymmetry of the PSF depend
significantly on the off-axis distance ($\theta$). The PSF is narrow and nearly
circular in the inner $\theta$\,$\lesssim$\,5' but becomes rapidly broader and
more asymmetries at larger off-axis, as demonstrated in the bottom panel of
Fig. \ref{acis} where we show sources detected in our data at four different
off-axis angles.

Extraction from circular regions containing a large fraction (e.g. 99\%) of the
PSF would guarantee to collect almost all source photons. However, given the
extended wings of the PSF, very large regions would be needed, incurring in the
risk of contamination from nearby sources, especially in crowded fields like
ours. Moreover the resulting inclusion of a large number of background events
would reduce the signal to noise of weak sources. On the other hand, extraction
from regions that are too small may reduce the photon statistic for further
spectral and timing analysis.

In order to tackle this problem, AE begins with calculating the shape of the
model PSF at each source position using the CIAO task {\sc mkpsf}. It then
refines the initial source positions (in our case estimated by PWDetect
assuming a symmetric PSF) by correlating the source images with the model
PSFs.  Following AE science 
hints\footnote{http://www.astro.psu.edu/xray/docs/TARA/ae\_users\_guide/node35},
this last procedure was only used for those sources lying at off-axis larger
than 5 arcmin (464 sources), while for the rest of source (539 sources) we
simply adopt data-mean positions. Coordinates listed in Table \ref{AEphot}, as
well as their 1$\sigma$ uncertainties, are the result of this process. We
verified that these positions are an improvement over those computed by
PWDetect, especially at large off-axis angles, comparing the offsets between
X-ray sources and counterparts in the 2MASS catalog (see \S 4.2). 

After re-computing positions, AE defines source extraction regions as polygonal
contours of the model PSF containing a specified fraction of source events
(f$_{\rm PSF}$). Generally, we chose f$_{\rm PSF}$=90\%, and computed the
contours from the PSF for a mono-energetic source with E\,=\,1.49 keV. For
$\sim$8\% of the sources in the denser parts of the \cyg field this fraction
was reduced so to avoid contamination with other nearby sources, in the most
extreme cases down to f$_{\rm PSF}$$\sim$40\%. The four panels at the bottom of
Fig. \ref{acis} show examples of extraction regions. Otherwise, we have
detected only six sources: Id.\# 17, 60, 488, 544, 568, 729 that suffers of a
pile-up fraction  (f$_{\rm pile-up}$) of 3.96\%, 17.7\%, 11.6\%, 2.39\%, 15.6\%
and 4.6\%, respectively. In these cases, we extracted events in annular regions
that exclude the PSF cores for all sources that exhibits a f$_{\rm pile-up}$
greater than 2\%. The inner radii was fixed at 2
arcsec\footnote{http://www.astro.msfc.nasa.gov/Ch4/Ch4\_15-03\_Tsujimoto.pdf},
while the outer circles were chooses as the smallest that inscribe the f$_{\rm
PSF}$=99\% contours.

Although the ACIS-I instrumental background level is spatially quite uniform,
the actual observed background varies substantially across the crowded \cyg
field due to the extended PSF wings of bright sources and  to their readout
trails. Background was therefore estimated locally for each source, adopting
once again the automated procedure implemented in AE, which defines background
extraction regions as circular annuli with inner radii 1.1 times the maximum
distance between the source and the 99\% PSF contour, and outer radii defined
so that the regions contains more than 100 ``background'' events. In order to
exclude contamination of the regions by nearby sources, background events are
defined from a ``swiss cheese'' image that excludes events within the inner
annuli radii of all the 1003 sources.

Results of the photon extraction procedure are listed in columns 8-16 of Table
\ref{AEphot}. We give: the source extraction area (column 8); the PSF fraction
within the extraction area, assuming E=1.49 keV (9); the background-corrected
extracted source counts in the 0.5-8.0 keV band (10); the exposure time (11),
the count rates (CR) in four spectral bands computed as the ratio between the
source photons (corrected for $f_{PSF}$) and the exposure time (12-15); the
median photon energy ($\overline {\rm E_x}$), (16).

In summary, our 1003 X-ray detections span a wide count range, from 4 to
$\sim$15000 photons. Most sources are faint (e.g. 42\% have less than 20
photons). The lower envelope of the counts vs. off-axis plot (not shown)
indicates that the minimum number of photons necessary for detection ranges
from 4 on axis to $\sim$20 close to the detector corners ($\theta$=10 
arcmin).

\begin{sidewaystable*}
\caption{Cygnus\,OB2 X-ray source catalog: first 35 rows. The complete 
table, containing 1003 rows is available in the electronic edition of A\&A.}
\label{AEphot}
\begin{tabular}{llllllllllllcccll}
\hline \hline \multicolumn{1}{l}{N$_{\rm x}$} &
\multicolumn{1}{l}{NAME} &
\multicolumn{1}{l}{R.A} &
\multicolumn{1}{l}{DEC.} &
\multicolumn{1}{l}{Error} &
\multicolumn{1}{l}{$\theta$} &
\multicolumn{1}{l}{Sig.} &
\multicolumn{1}{l}{Area} &
\multicolumn{1}{l}{PSF} &
\multicolumn{1}{l}{Cts} &
\multicolumn{1}{l}{Exp. Time} &
\multicolumn{4}{l}{Count Rates ($\times$10$^{-3}$ ctn\,s$^{-1}$)} &
\multicolumn{1}{l}{$\overline E_{\rm x}$} &
\multicolumn{1}{l}{Var.} \\
\cline{12-15}
\multicolumn{1}{l}{\#} &
\multicolumn{1}{l}{CXOCYG\,J+} &
\multicolumn{1}{l}{[h:m:s]} &
\multicolumn{1}{l}{[d:m:s]} &
\multicolumn{1}{l}{(")} &
\multicolumn{1}{l}{(')}&
\multicolumn{1}{l}{($\sigma$)} &
\multicolumn{1}{l}{(px.)} &
\multicolumn{1}{l}{(\%)} &
\multicolumn{1}{l}{(ph.)}&
\multicolumn{1}{l}{(ks)} &
\multicolumn{1}{l}{Tot.}&
\multicolumn{1}{c}{Soft}&
\multicolumn{1}{c}{Med.}&
\multicolumn{1}{c}{Hard}&
\multicolumn{1}{l}{(keV)}&
\multicolumn{1}{l}{log(P$_{\rm ks}$)} \\
\hline
   1 & 203225.09+411019.6 & 20:32:25.09 & 41:10:19.60 &  0.62 & 10.03 &   6.55 &   1170 & 0.90 &    67 & 87.8255 &   0.855 &   0.135 &   0.233 &   0.487 & 3.23 & -4.00 $\dag$\\
   2 & 203225.55+410847.3 & 20:32:25.55 & 41:08:47.36 &  0.70 & 10.76 &   6.66 &   1429 & 0.89 &    72 & 92.9360 &   0.871 &   0.620 &   0.059 &   0.192 & 1.35 & -0.65  \\
   3 & 203225.97+411054.7 & 20:32:25.97 & 41:10:54.77 &  0.75 &  9.62 &   4.28 &    973 & 0.90 &    35 & 90.9414 &   0.434 &   0.099 &   0.153 &   0.182 & 2.37 & -0.32  \\
   4 & 203227.51+411358.3 & 20:32:27.52 & 41:13:58.37 &  0.88 &  8.47 &   2.30 &    630 & 0.89 &    14 & 92.5496 &   0.173 &   0.004 &   0.060 &   0.109 & 3.18 & -0.17  \\
   5 & 203227.62+410831.4 & 20:32:27.63 & 41:08:31.46 &  0.76 & 10.61 &   5.92 &   1420 & 0.91 &    55 & 96.8402 &   0.637 &   0.160 &   0.198 &   0.278 & 2.36 & -0.00  \\
   6 & 203227.70+411317.0 & 20:32:27.71 & 41:13:17.07 &  0.26 &  8.55 &  13.38 &    618 & 0.89 &   216 & 96.5481 &   2.521 &   0.529 &   0.862 &   1.130 & 2.61 & -4.00 $\dag$\\
   7 & 203228.88+410807.5 & 20:32:28.89 & 41:08:07.59 &  0.80 & 10.67 &   5.09 &   1623 & 0.90 &    49 & 96.6863 &   0.571 &   0.136 &   0.167 &   0.268 & 2.75 & -2.37 $\dag$\\
   8 & 203229.12+411400.7 & 20:32:29.12 & 41:14:00.78 &  0.55 &  8.16 &   4.09 &    545 & 0.89 &    33 & 92.8669 &   0.400 &   0.130 &   0.078 &   0.192 & 2.62 & -0.59  \\
   9 & 203229.26+410850.1 & 20:32:29.26 & 41:08:50.17 &  0.59 & 10.17 &   6.70 &    834 & 0.82 &    64 & 92.6690 &   0.842 &   0.364 &   0.202 &   0.276 & 2.02 & -0.44  \\
  10 & 203229.84+411454.2 & 20:32:29.85 & 41:14:54.27 &  0.50 &  7.97 &   5.56 &    525 & 0.90 &    50 & 93.3914 &   0.605 &   0.249 &   0.185 &   0.171 & 2.24 & -0.67  \\
  11 & 203229.91+411056.0 & 20:32:29.92 & 41:10:56.06 &  0.72 &  8.94 &   2.77 &    751 & 0.90 &    19 & 92.9014 &   0.227 &   0.080 &   0.058 &   0.089 & 2.30 & -0.97  \\
  12 & 203230.63+410829.1 & 20:32:30.64 & 41:08:29.11 &  0.58 & 10.19 &   8.52 &   1205 & 0.90 &    99 & 94.5473 &   1.173 &   0.320 &   0.478 &   0.375 & 2.14 & -0.08  \\
  13 & 203230.63+410854.8 & 20:32:30.63 & 41:08:54.82 &  0.66 &  9.92 &   5.21 &    784 & 0.83 &    44 & 96.4507 &   0.552 &   0.067 &   0.180 &   0.305 & 2.93 & -1.14 $\dag$\\
  14 & 203230.88+411029.5 & 20:32:30.89 & 41:10:29.53 &  0.46 &  8.99 &   7.08 &    761 & 0.90 &    75 & 96.5167 &   0.867 &   0.063 &   0.176 &   0.628 & 3.42 & -0.07  \\
  15 & 203231.39+410955.9 & 20:32:31.40 & 41:09:55.98 &  0.57 &  9.21 &   6.16 &    821 & 0.90 &    60 & 94.7546 &   0.714 &   0.259 &   0.168 &   0.287 & 2.27 & -0.01  \\
  16 & 203231.42+411335.0 & 20:32:31.43 & 41:13:35.01 &  0.55 &  7.80 &   3.85 &    480 & 0.90 &    30 & 93.4071 &   0.357 &   0.039 &   0.098 &   0.220 & 3.01 & -0.33  \\
  17 & 203231.53+411407.9 & 20:32:31.53 & 41:14:07.98 &  0.05 &  7.70 &  55.14 &    431 & 0.89 &  3173 & 97.6726 &  36.534 &  14.545 &  13.742 &   8.247 & 1.85 & -2.83  \\
  18 & 203231.58+411712.5 & 20:32:31.58 & 41:17:12.52 &  0.73 &  7.95 &   2.01 &    517 & 0.90 &    12 & 92.8637 &   0.145 &   0.007 &   0.049 &   0.089 & 3.10 & -0.04  \\
  19 & 203232.32+411507.3 & 20:32:32.32 & 41:15:07.30 &  0.56 &  7.50 &   3.55 &    418 & 0.89 &    25 & 94.0133 &   0.306 &   0.031 &   0.097 &   0.177 & 3.23 & -0.02  \\
  20 & 203232.44+411312.3 & 20:32:32.45 & 41:13:12.36 &  0.43 &  7.69 &   5.60 &    413 & 0.89 &    49 & 93.2124 &   0.592 &   0.112 &   0.202 &   0.278 & 2.64 & -2.69 $\dag$\\
  21 & 203233.28+411058.8 & 20:32:33.29 & 41:10:58.88 &  0.75 &  8.36 &   2.82 &    575 & 0.90 &    18 & 88.3312 &   0.230 &   0.162 &   0.034 &   0.034 & 1.15 & -0.44  \\
  22 & 203233.47+411218.0 & 20:32:33.48 & 41:12:18.04 &  0.49 &  7.77 &   4.49 &    459 & 0.90 &    34 & 91.7675 &   0.415 &   0.177 &   0.127 &   0.111 & 1.82 & -0.04  \\
  23 & 203233.53+411405.7 & 20:32:33.54 & 41:14:05.74 &  0.38 &  7.33 &   5.47 &    364 & 0.90 &    50 & 95.7566 &   0.593 &   0.199 &   0.201 &   0.193 & 1.92 & -0.49  \\
  24 & 203234.33+411506.3 & 20:32:34.34 & 41:15:06.33 &  0.54 &  7.12 &   3.71 &    350 & 0.90 &    26 & 94.3211 &   0.311 &   0.074 &   0.056 &   0.181 & 3.52 & -0.44  \\
  25 & 203234.45+411154.7 & 20:32:34.45 & 41:11:54.74 &  0.58 &  7.75 &   3.96 &    446 & 0.90 &    27 & 94.8080 &   0.326 &   0.061 &   0.203 &   0.062 & 2.06 & -0.26  \\
  26 & 203235.17+411143.4 & 20:32:35.18 & 41:11:43.48 &  0.47 &  7.70 &   4.97 &    430 & 0.90 &    39 & 91.6010 &   0.475 &   0.196 &   0.159 &   0.120 & 1.96 & -0.03  \\
  27 & 203235.65+411509.5 & 20:32:35.65 & 41:15:09.50 &  0.33 &  6.88 &   7.09 &    312 & 0.90 &    70 & 96.3691 &   0.815 &   0.228 &   0.368 &   0.219 & 2.10 & -1.17  \\
  28 & 203235.86+411248.4 & 20:32:35.87 & 41:12:48.42 &  0.37 &  7.18 &   5.22 &    340 & 0.90 &    43 & 93.8249 &   0.519 &   0.083 &   0.289 &   0.147 & 2.04 & -2.14 $\dag$\\
  29 & 203235.96+412135.6 & 20:32:35.97 & 41:21:35.64 &  0.78 &  9.47 &   4.51 &    669 & 0.83 &    33 & 95.8351 &   0.413 &   0.190 &   0.064 &   0.159 & 1.85 & -0.11  \\
  30 & 203236.32+411900.7 & 20:32:36.32 & 41:19:00.76 &  0.47 &  7.84 &   4.96 &    489 & 0.90 &    41 & 94.6415 &   0.484 &   0.209 &   0.129 &   0.146 & 1.76 & -0.44  \\
  31 & 203236.38+411831.0 & 20:32:36.39 & 41:18:31.08 &  0.62 &  7.59 &   3.81 &    403 & 0.89 &    26 & 90.0839 &   0.333 &   0.242 &   0.060 &   0.032 & 1.41 & -1.46  \\
  32 & 203236.50+411810.3 & 20:32:36.50 & 41:18:10.36 &  0.67 &  7.42 &   3.78 &    376 & 0.89 &    26 & 91.6356 &   0.320 &   0.115 &   0.112 &   0.093 & 1.90 & -1.55  \\
  33 & 203236.62+412214.5 & 20:32:36.63 & 41:22:14.52 &  0.26 &  9.85 &  14.02 &    689 & 0.81 &   232 & 97.5187 &   2.949 &   1.112 &   0.835 &   1.002 & 1.93 & -4.00 $\dag$\\
  34 & 203236.86+411944.6 & 20:32:36.86 & 41:19:44.62 &  0.40 &  8.16 &   7.22 &    544 & 0.90 &    74 & 97.0381 &   0.859 &   0.295 &   0.288 &   0.277 & 1.86 & -0.83  \\
  35 & 203236.84+412146.1 & 20:32:36.85 & 41:21:46.11 &  0.56 &  9.48 &   5.65 &    660 & 0.83 &    46 & 95.1661 &   0.592 &   0.278 &   0.162 &   0.153 & 1.82 & -1.24  \\
\hline
\end{tabular}
\smallskip

Notes -- $\dag$: flare-like sources.
\end{sidewaystable*}

\subsection{X-ray hardness ratios}
\label{sect:hardratio}
 
In order to characterize the X-ray spectra of low-statistic sources it is
common practice to use the ratios of source counts in different spectral bands,
i.e. X-ray hardness ratios (XHR), or the logarithm of these values which can be
considered ``X-ray colors'' \citep{1989A&A...225...48S,2003ApJ...595..719P}.

\begin{figure*}[!ht] \centering \includegraphics[width=18cm,angle=0]{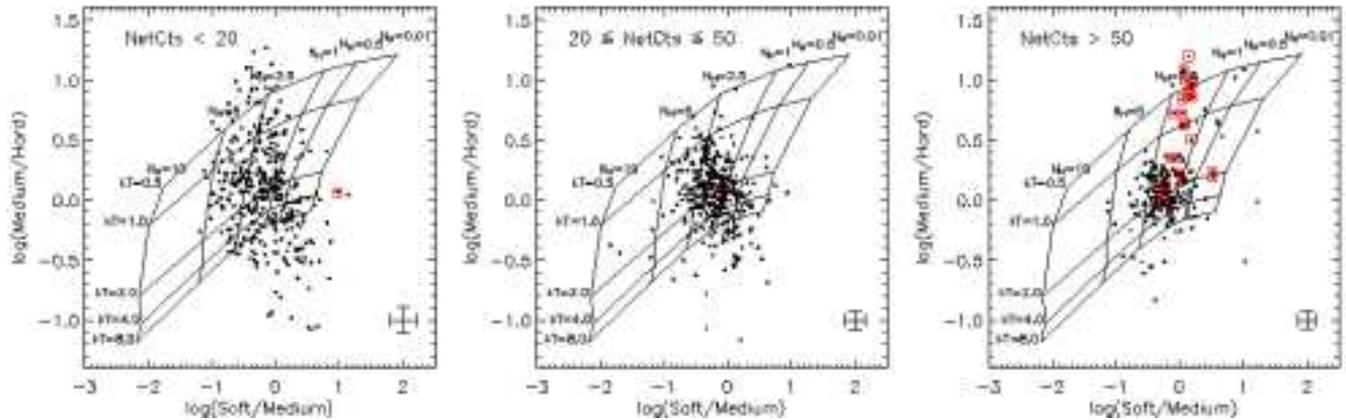}
\caption{X-ray color-color diagrams for our ACIS sources. The three panels
refer to sources in three ranges of detected counts, as indicated in the
upper-left corners. Colors are defined for three energy bands: {\em Soft}
(0.5-1.7 keV),  {\em Medium} (1.7-2.8 keV) and  {\em Hard} (2.8-8.0 keV) keV.
The grids refers to predicted colors for Raymond-Smith thermal emission models
with different temperatures (kT\,=\,0.5, 1.0, 2.0, 4.0, and 8.0 keV) and
affected by varying amounts of absorption, N$_{\rm H}$\,=\,0.01, 0.5, 1.0, 2.5,
5, and 10 $\times$10$^{22}$ cm$^{-2}$. Boxes refer to known O- and early B-type
stars. Note that the dispersion of points is larger for low-statistics sources.
Typical 1$\sigma$ error bars are plotted in the
right-bottom part of each panel.} \label{grid-model} \end{figure*}

We use this method, dividing the full energy range into three bands: Soft
(S$_{\rm x}$: 0.5-1.7 keV), Medium (M$_{\rm x}$: 1.7-2.8 keV) and Hard (H$_{\rm
x}$: 2.8-8.0 keV). Figure \ref{grid-model} shows, separately for sources in
three detected counts ranges, the ``Hard X-ray color'' ($\log [M_{\rm x}/H_{\rm
x}]$) vs. the ``Soft X-ray color'' ($\log [S_{\rm x}/M_{\rm x}]$). For
reference we also plot the predicted loci for absorbed thermal sources with
plasma temperatures between 0.5 and 8.0 keV and  N$_{\rm H}$ between 10$^{20}$
and 10$^{23}$ cm$^{-2}$. The grid was calculated with the {\it Portable
Interactive Multi-Mission Simulator} 
(PIMMS\footnote{http://heasarc.gsfc.nasa.gov/docs/software/tools/pimms.html})
for a Raymond-Smith (RS) emission model \citep{1977ApJS...35..419R}.

A comparison of the three panels in Fig. \ref{grid-model} shows that the
position of sources with respect to the kT-N$_{\rm H}$ grid is significantly
affected, other than by the source spectra, by the source statistic. The
positions of individual sources with less than 20 counts (left panel) are, for
example, considerably spread-out because of large statistical uncertainties. 
Sources with higher statistics (20\,$\leq$\,NetCnts\,$\leq$\,50, central panel)
are more concentrated around kT\,$\sim$\,2.0 keV and N$_{\rm
H}$\,$\sim$\,2.3$\times$10$^{22}$  cm$^{-2}$\,. Sources with more than 50
photons (right panel) are even  more concentrated around these same values of
kT and N$_{\rm H}$.  A separate group of soft-spectrum sources becomes however
visible  in the upper part of the grid, with typical kT$\sim$0.8 keV.  Most of
these sources are associated with known early type stars, indicated by squares
in Fig. \ref{grid-model}. Others, with somewhat smaller N$_{\rm H}$, are likely
foreground stars as also argued in \S \ref{sect:NIRprop}.

\section{Optical and Near-IR counterparts}

\cyg has been the target of numerous optical studies for more than 50 years
\citep{1954ApJ...119..344J,1958ApJ...128...41S}. \citet{1991AJ....101.1408M}
identified 120 candidate massive members in an area of about 0.35 deg$^2$, on
the basis of UBV photometry, and   gave optical spectral classifications for
over 70 OB stars. More recently, \citet{2003ApJ...597..957H} published new
spectral classification for 14 more OB candidates, improving the massive star
census of Cygnus OB2, and the estimates for its distance (1450\,pc) and age
($\sim$2\,Myr). A total of 33 OB stars from the catalog of
\citet{2003ApJ...597..957H} lie in the 0.0823 deg$^2$ FOV of our X-ray
observation and were thus cross identified with our X-ray source list (\S 3.2).
Due to the high and variable absorption in the \cyg line of sight, and to the
relative shallowness of the available optical catalogs, the correlation between
X-ray sources and optical catalogs is of limited use for the study of
intermediate- and low-mass stars. We thus decided to base our characterization
of the X-ray sources mostly on near-IR data, for which the impact of dust
extinction is reduced and actually comparable to that on the X-ray band. We
made use of J (1.25 $\mu$m), H (1.65 $\mu$m) and K$_{\rm s}$ (2.17 $\mu$m)
photometry from the {\it Two Micron All Sky Survey} (2MASS) Point Source
Catalog (PSC)\footnote{See http://www.ipac.caltech.edu/2mass}. 2MASS is
complete to magnitudes of 15.8, 15.1 and 14.3 mag in the J, H and K$_s$ bands,
respectively. We limited our analysis to 2MASS sources for which the quality
flag for at least one of the three magnitudes is equal to A, B, C or D (cf. the
2MASS All-Sky Data Release User's Guide). With this restriction 11 sources were
removed from our initial list of 5061 2MASS point sources in the ACIS FOV,
leaving a total of 5050 objects. 

\subsection{Cross-identifications}

We began by cross identifying our X-ray source list with the 2MASS catalog.
Identification radii, R$_{\rm id}$, were chosen so to limit the number of
spurious identifications due to chance alignments, N$_{\rm chance}$, and at the
same time to include a large number of the true physical associations, N$_{\rm
true}$. First we estimated, as a function of  R$_{\rm id}$, the number of
chance identifications expected in a given sky area, A$_{\rm search}$, assuming
(i) uncorrelated NIR and X-ray positions and (ii) a uniform surface source
density of the N$_{\rm 2MASS}$ 2MASS sources lying in the search area: N$_{\rm
chance}$(R$_{\rm id}$)=N$_{\rm X}$A$_{\rm id}$N$_{\rm 2MASS}$/A$_{\rm search}$,
where A$_{\rm id}$=$\pi$R$_{\rm id}^2$ is the area of identification circles.
N$_{\rm true}$(R$_{\rm id}$) can instead be estimated from the observed total
number of identifications at any given R$_{\rm id}$ as N$_{\rm id}$(R$_{\rm
id}$)-N$_{\rm chance}$(R$_{\rm id}$). We chose the identification radius as the
largest for which N$_{\rm true}>$N$_{\rm chance}$.

Because the \chandra PSF, and therefore the position uncertainty of X-ray
sources, depends mainly on the off-axis, we perform this analysis in four
different ranges of off-axis: [0-2), [2-4), [4-7) and $>$ 7 arcmin. Our
identification radii in these four regions are 1.0, 1.5, 2.1, and 2.7 arcsec,
respectively. 

Before performing the final identifications we searched for possible systematic
differences between the X-ray and 2MASS positions. We first performed a
preliminary cross-identification and compared the coordinates of identified
pairs. A small systematic offset between the two catalogs was found. We thus
shifted the \chandra coordinates and performed a new cross-identification,
repeating this process iteratively until the offset was reduced to $\sim$\,0.01
arcsec, i.e. much smaller than the statistical errors on source positions. In
the end the offset between the two catalogs was: $\Delta$(RA$_{\rm
x-2MASS}$)\,=\,+0.12"$\pm$0.10"  and $\Delta$(DEC$_{\rm
x-2MASS}$)\,=\,+0.49"$\pm$0.17". The result of the final identification is
shown in Table \ref{identif}, where we lists, in the first 7 columns, the X-ray
and near-IR identifiers of cross-identified sources, the offset between the two
positions, and the J, H, K$_s$ magnitudes from 2MASS.  A total of 775 X-ray
sources out of the 1003 in our list were identified  with 2MASS objects. Three
X-ray sources (\#12, \#148, and \#449) were identified with two 2MASS objects
but, after checking the positions visually, we adopted the closest of the
counterparts. The fraction of identified X-ray sources appears to  increase as
we move to larger off-axis. In the four annular regions defined  above these
fractions are: 76/116 (65\%), 227/304 (74\%), 294/374 (79\%) and 178/209
(85\%), in order of increasing off-axis. This trend can be attributed to the
already mentioned (\ref{sect:extraction}) dependence of the ACIS-I sensitivity
with off-axis angle and to the fact that more intense X-ray sources are more
likely to have a near-IR counterpart than fainter ones.

With respect to the expected number of chance identifications we estimate, for
off-axis ranges [0:2), [2:4), [4:7), and $>$7 arcmin, and with the formula
given above, no more than 1.6, 7.9, 16.4, and 12.5, respectively for a total of
$\leq$ 39. Note however that the assumption that X-ray and 2MASS source lists
are fully uncorrelated is quite certainly not true (other than for the 10
expected spurious detections), so that this number can actually only be
considered as a loose upper limit \citep[cf.][]{2003ApJ...588.1009D}.

We estimate the expected number of extragalactic sources in our detection list
following \cite{2006astro.ph..4243F}. We consider the ACIS count-rates of
non-stellar sources in the {\em Chandra Deep Field North}
\citep[CDFN,][]{2003AJ....126..539A,2003AJ....126..632B} and estimate
absorption corrected count-rates assuming N$_{\rm H}$\,=\,1.54$\times10^{22}$ 
(from \Av = 7.0, see \S \ref{sect:NIRprop}) using PIMMS and assuming power-law
spectra with index  $\Gamma$. We then compare these count rates with upper
limits taken at random position in the ACIS FOV. For $\Gamma$ between 1 and 2
we obtain 61 to 87 expected extragalactic sources. Given the intrinsic near-IR
fluxes of these sources and the absorption toward Cygnus OB2, they are expected
to be among the 228 without NIR counterparts
\citep[c.f.][]{2006astro.ph..4243F}.

Finally we identified our X-ray source list with the 33 early type stars listed
by \citet{2003ApJ...597..957H} and lying in our ACIS FOV. Of these, all the 20
O-type stars result associated with an X-ray source, while of the 13 B-type
stars only 6 are detected. 

{\small
\begin{table*}
\caption{Near-IR counterparts of Cygnus\,OB2 X-ray sources and near-IR stellar parameters: first rows. 
The complete version is available in the electronic edition of A\&A. }
\label{identif}
\begin{tabular}{cccclllcc}
\multicolumn{9}{c}%
{{\bfseries}}\\
\hline \hline
\multicolumn{1}{l}{N$_{\rm x}$} &
\multicolumn{2}{c}{X-ray - 2MASS counterpart} &
\multicolumn{1}{l}{Off.} &
\multicolumn{4}{c}{2MASS photometry} &
\multicolumn{1}{l}{A$_{\rm v}^\ddag$} \\
\cline{2-3} \cline{5-8}
\multicolumn{1}{l}{\#} &
\multicolumn{1}{c}{CXOAC\,J+} &
\multicolumn{1}{c}{2MASS\,J+} &
\multicolumn{1}{l}{(")} &
\multicolumn{1}{l}{J}&
\multicolumn{1}{l}{H}&
\multicolumn{1}{l}{K$_{\rm s}$}&
\multicolumn{1}{l}{Ph.\,qual$^\dag$} &
\multicolumn{1}{l}{(mag)} \\
\hline
   1 & 203225.09+411019.6 &       $--------$ & $--$ & $-----$ & $-----$ & $-----$ & $---$ & $--$ \\
   2 & 203225.55+410847.3 & 20322545+4108473 & 1.56 & 14.76$\pm$0.06 & 14.23$\pm$0.07 & 13.95$\pm$0.07 &  AAA &  8.54 \\
   3 & 203225.97+411054.7 & 20322597+4110547 & 0.24 & 15.24$\pm$0.05 & 13.65$\pm$0.03 & 12.69$\pm$0.03 &  AAA &  7.71 \\
   4 & 203227.51+411358.3 & 20322756+4114001 & 1.61 & 16.00$\pm$0.06 & 14.46$\pm$0.04 & 13.90$\pm$0.06 &  AAA &  0.16 \\
   5 & 203227.62+410831.4 & 20322754+4108336 & 2.41 & 15.91$\pm$0.06 & 14.39$\pm$0.03 & 13.82$\pm$0.05 &  AAA &   N/A \\
   6 & 203227.70+411317.0 & 20322768+4113169 & 0.58 & 14.96$\pm$0.03 & 13.49$\pm$0.02 & 12.88$\pm$0.03 &  AAA &  2.21 \\
   7 & 203228.88+410807.5 &       $--------$ & $--$ & $-----$ & $-----$ & $-----$ & $---$ & $--$ \\
   8 & 203229.12+411400.7 & 20322913+4114012 & 0.26 & 16.74$\pm$0.16 & 15.36$\pm$0.08 & 14.72$\pm$0.11 &  CAB &  9.01 \\
   9 & 203229.26+410850.1 & 20322928+4108494 & 1.03 & 12.26$\pm$0.01 & 11.49$\pm$0.01 & 11.07$\pm$0.01 &  AAA &  8.52 \\
  10 & 203229.84+411454.2 & 20322985+4114536 & 0.83 & 17.44          & 15.43$\pm$0.10 & 14.37$\pm$0.09 &  UAA &   N/A \\
  11 & 203229.91+411056.0 & 20322994+4110575 & 1.29 & 15.26$\pm$0.05 & 14.04$\pm$0.04 & 13.53$\pm$0.04 &  AAA &   N/A \\
  12 & 203230.63+410829.1 & 20323051+4108284 & 2.19 & 15.47$\pm$0.07 & 13.71          & 13.09          &  AUU &  4.19 \\
  13 & 203230.63+410854.8 & 20323064+4108565 & 1.49 & 15.31$\pm$0.05 & 13.68$\pm$0.03 & 13.01$\pm$0.03 &  AAA &  5.43 \\
  14 & 203230.88+411029.5 &       $--------$ & $--$ & $-----$ & $-----$ & $-----$ & $---$ & $--$ \\
  15 & 203231.39+410955.9 & 20323143+4109558 & 0.58 & 14.52$\pm$0.05 & 13.04 & 12.46 &  AUU &  9.69 \\
\hline
\end{tabular}
\smallskip

\dag 2MASS photometric quality flags for the J, H and, K$_s$ bands: ``A'' to
``D'' indicate decreasing quality of the measurements, ``U'' that the value is
an upper limit. See 2MASS documentation for details.  \ddag N/A indicates stars
with unconstrained \Av (see text). 
\end{table*} }

\subsection{Near-IR properties of the X-ray sources}
\label{sect:NIRprop}

We now investigate the NIR properties of the X-ray sources with a 2MASS
counterparts. For this purpose we restrict our analysis to  sources with high
quality photometry, i.e. those for which the {\em quality flag} (see 2MASS
documentation) is '$AAA$', or for which uncertainties on the J, H and K$_s$ 
magnitudes are all lower than 0.1 mag. With these requirements the total number
of IR sources in the ACIS FOV is reduced from 5050 to 2187. Counterparts of
X-ray sources were selected only on the basis of their magnitude errors ($<$0.1
mag), yielding 519 sources out of the original 775. 

\begin{figure}[!h] \centering
\includegraphics[width=8.5cm,angle=0]{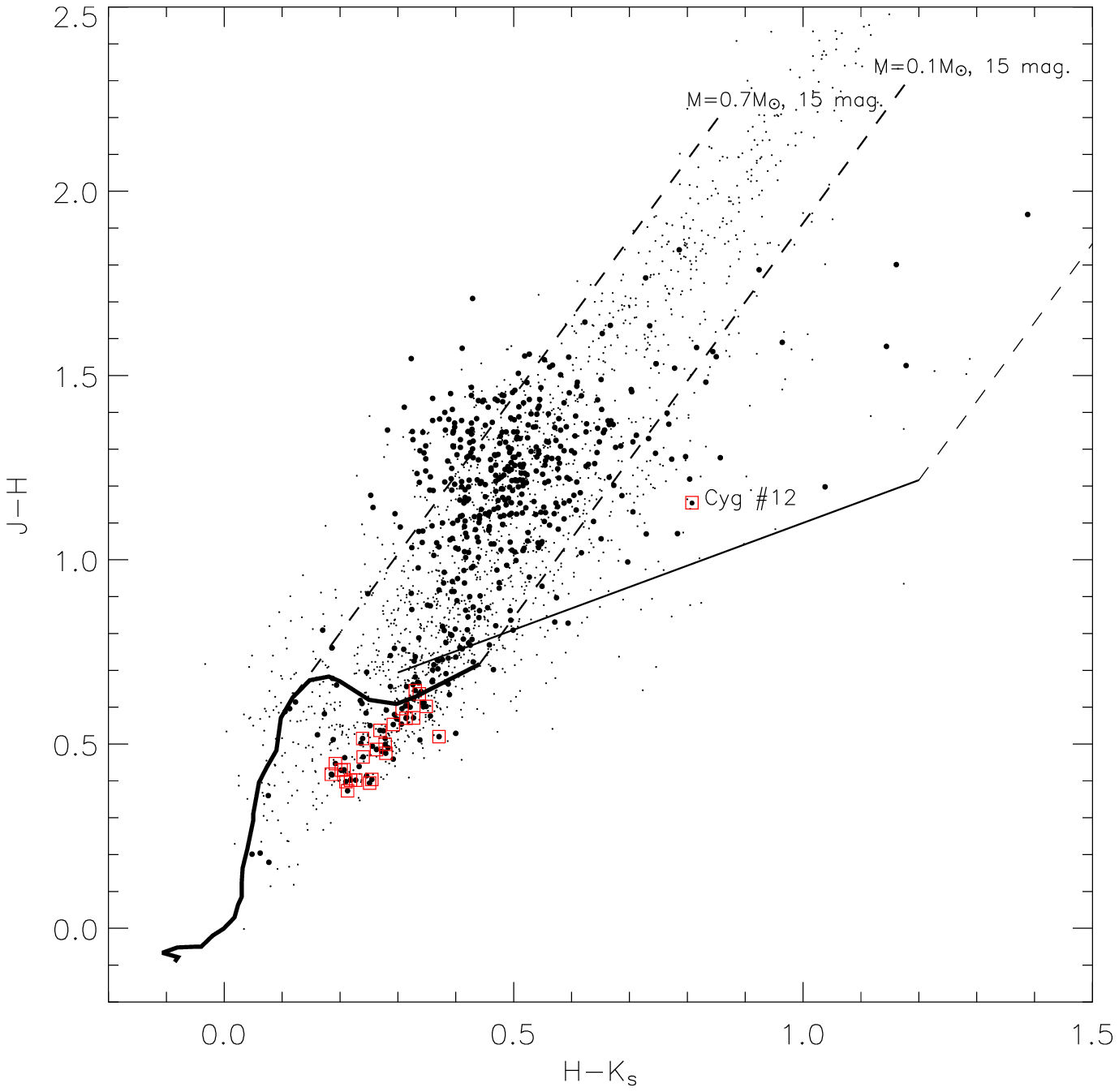} \caption{JHK$_{s}$
color-color diagram of objects in the  17'$\times$17'\cyg FOV with high quality
2MASS photometry. Filled circles and small points refer to X-ray detected and
undetected 2MASS sources, respectively. Detected known O- and B-type stars are
indicated by squares. We also show for reference the main-sequence from 
\cite{1995ApJS..101..117K}, the CTTS locus of  \citet{1997AJ....114..288M} and
reddening vectors ({\it dashed lines}) with length corresponding to \Av\,=\,15
mag. Note: i) the low number of X-ray sources lying in the CTTS loci; ii) the
peculiar position of Cyg\,\#12 (labeled), one of most intrinsically luminous
stars in the Galaxy, indicative of a high reddening.} \label{jh_hk}
\end{figure}

Figure \ref{jh_hk} shows the J-H vs. H-K$_s$ color-color (CC) diagram for all
the selected 2MASS objects in the ACIS FOV, both X-ray detected and undetected.
We also plot for comparison the MS  \citep{1995ApJS..101..117K}, the Classical
T-Tauri Stars (CTTS) locus of \citet{1997AJ....114..288M}, and three reddening
vectors starting from these loci and with slope ($\rm A_{Ks}/E(H-K_s)$=0.125)
taken from the extinction law given by \cite{2003ApJ...597..957H}. \cyg members
with purely photospheric emission should lie in this reddening band. However,
as the line of sight toward \cyg is  not far from that to the Galactic center,
many field interlopers are also expected in the same band. Young stellar
objects (YSOs), such as Classical T\,Tauri and Herbig Ae/Be stars, because of
the NIR excess emission originated in the inner parts of their circumstellar
disks, are often found to the right of this band, i.e. in the CTTS locus.

We note here, however, that very few X-ray sources, i.e. likely \cyg members,
have colors consistent with the (reddened) CTTS locus.  If we neglect the
uncertainties on data points, 23 sources lie in the reddening band of the CTTS
locus. We do not include here the peculiar supergiants B5\,Ie Cyg\#12 and
discuss its nature in \S\ref{sect:reshi}. A total of 23 stars in the CTTS
reddening band means a fraction of  23/519$\sim$\,4.4\% with respect to all the
X-ray sources in the CC diagram. We compare this fraction with the one observed
in the ONC, adopting Chandra Orion Ultradeep Project (COUP) sources in the same
mass range\footnote{We only consider COUP sources for which mass estimates are
given by \citet{2005ApJS..160..353G}} that we reach in \cyg (i.e. M$\geq$\,1
M$_\odot$, c.f. \S \ref{sect:results}): 19/92$\sim$\,20.6\%. The difference
between the two fractions of stars with near-IR detected disks is significant,
a factor of $\sim$4.7. It could be due to two different reasons:

- \cyg is older than the ONC ($\sim$\,2 vs. $\sim$1\,Myrs) and its disks might
have dissipated. According to \cite{2005astro.ph.11083H}, the fraction of stars
with disks detectable in the near-IR should decrease between 1 and 2 Myr by a
factor of 2-3. This, together with the statistical uncertainties in the disk
fractions might explain the observed difference.\footnote{A direct comparison
with the results of \cite{2005astro.ph.11083H} is however not possible because
her estimate of disk lifetime (i) refers to star in the 0.3-1.0 M$_\odot$, i.e.
less massive than the ones we observe in Cygnus OB2, (ii) is based on a
different, and more efficient, indicator of disks presence than the one we can
use here, i.e. the H-K color excess measured with respect to the photospheric
value as determined from spectral types.}

- The effect of disk photo-evaporation by the intense UV radiation field of the
hot massive OB stars. This mechanism, predicted theoretically
\citep[e.g.][]{2004ApJ...611..360A} and observed, e.g., in the ONC proplyds
\citep[e.g.][]{2000AJ....119.2919B}, is thought to be effective in massive star
forming regions, such as Cygnus OB2, having a large number of UV emitting O-
and B-type stars. A shortening of the disk lifetime with decreasing distance
from O- stars has been recently observed in the massive star forming region 
NGC\,6611 \citep[][in prep.]{Guarcello_in_press}.

\begin{figure}[ht] \includegraphics[width=8.9cm,angle=0]{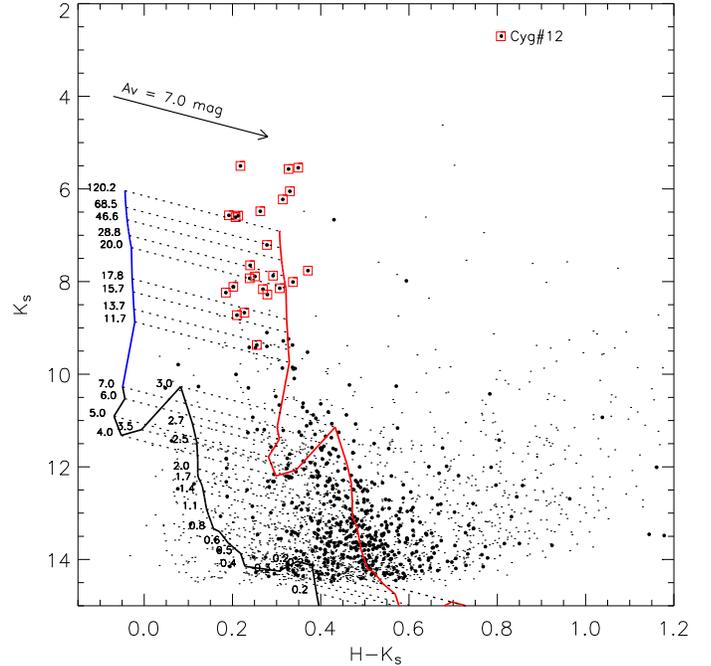}
\caption{CM diagrams of the \cyg region. Symbols as in figure \ref{jh_hk}. The
two parallel curves indicate the expected cluster loci for the assumed distance
and for a mean reddening of A$_V$=0.0 and 7.0 mag. They represent the MS for
(un-reddened) K$_{\rm s0}<$10.3 and the 2\,Myr isochrone for fainter
magnitudes.} \label{k_hk} \end{figure}

Figure \ref{k_hk} shows the K$_{\rm s}$ vs. H-K$_{\rm s}$ color magnitude (CM)
diagram for the same stars plotted in Fig. \ref{jh_hk}. We also show for
reference the expected cluster locus: the intrinsic K$_{\rm s}$ magnitudes and
H-K$_{\rm s}$ colors for stars earlier than B5V were taken from the MS
calibration of  \citet{2000A&A...360..539K} and \citet{1989LNP...341...61B},
respectively. For later spectral types (masses between 0.1 and 7 M$_\odot$), we
adopted the 2Myr isochrone from \cite{2000A&A...358..593S}, converted to the
observational plane using the calibration given by \citet{1995ApJS..101..117K}.
The adopted MS and 2Myr isochrone overlap satisfactorily.

In order to estimate the typical visual absorption of cluster members we
calculated the distance of each X-ray source to the cluster locus along the
reddening direction. Resulting \Av values for individual sources are listed in
the last column of table\,\ref{identif}. Note that for 10.25$<$K$_{\rm
s0}<$11.35 the absorption cannot be constrained because the reddening vector
intersect the cluster locus more than once. We computed median absorption
values in two luminosity ranges: for K$_{\rm s0}>$11.35 we obtained \Av=7.0
mag, while for K$_{\rm s0}<$10.25, considering only the known OB stars, we
obtained  \Av=5.63 mag. We note that the above estimations depend on the
reliability of the assumed cluster locus and on the assumption that the H and K
magnitudes are not significantly affected by disk-induced excesses. As observed
above, given the paucity of stars with excesses, the latter appears to be a
good approximation for \cyg stars. It is anyway comforting that our estimate
for the OB type stars is in fairly good agreement with the median of the
extinctions computed by  \cite{2003ApJ...597..957H} using photometric and
spectroscopic data: \Av=5.7 mag. The lower extinction derived for high mass
stars with respect to lower mass ones may indicate that the strong winds and
radiation field of massive stars may have cleared their surrounding
environment.

We note that in both the CC and CM diagrams, some X-ray sources, of the order
of 15-20, lie close to the un-reddened cluster loci. These are likely to be
foreground MS stars and thus to contaminate the sample of X-ray detected
cluster members. This conclusion is corroborated by the  X-ray hardness-ratio
analysis: eight of these stars have indeed soft and relatively unabsorbed
spectra (the stars to the right of the OB stars, in the upper-right corner of
Fig. \ref{grid-model}-right), as expected from foreground field stars.

In the following (\S \ref{sect:results}) we will correlate the X-ray properties
of our sources with stellar parameters derived from the available optical and
near-IR data. We obtain an estimate of stellar masses for 682 2MASS
counterparts from the J-band magnitudes (limited to {\em quality flags} ``A''
to ``D'') and the mass vs. J mag. relationship appropriate to the cluster age,
distance and extinction, obtained as described above for the cluster locus in
the H, H-K diagram.\footnote{The use of the J-band is justified because (i) in
the presence of disk excesses the J-band is the most representative of the
photospheric emission and (ii) the mass ranges in which is the mass-luminosity
relationship is degenerate are narrower than for similar relationship with the
H and K bands.} The relation is degenerate in two ranges of J, corresponding to
0.2-0.4M$_\odot$ and 2.8-5.3M$_\odot$. Four and 28 X-ray detected stars lie in
these ranges, respectively.

\section{Temporal variability}
\label{sect:var}

The temporal variability of young low mass sources is often complex: the most
common phenomena are magnetic flares with rapid rise and slower decays,
superposed on an apparently constant  or, sometimes, rotationally modulated
emission \citep{2005ApJS..160..450F, 2005ApJS..160..423W}.  Other forms of
variability, with less clear physical origin, are however often observed.

We first investigated X-ray variability using the non-binned one-sample
Kolmogorov-Smirnov (KS) test \citep{1992nrfa.book.....P}. This test compares
the distribution of photon arrival times with that expected for a constant
source. The test was applied to photons in the source extraction regions, which
also contain background photons. Given that the background was found to be low
and constant (\S 2.1), the results, i.e. the confidence with which we can
reject the hypothesis that the flux was constant during our observation, can be
attributed to the source photons. Table \ref{AEphot}, column 17, reports the
logarithm of the KS-test significance with values $<$\,-4 truncated at that
value: sources with log(P$_{\rm KS}$)\,$<$\,-3.0 can be considered almost
definitively variable as we expect at most one of the 1003 sources (i.e. 0.1\%)
to be erroneously classified as variable. Eighty-five X-ray sources 
($\approx$8.5\% of the total) fall in this category. Sources with
-2.0\,$<$\,log(P$_{\rm KS}$)\,$<$\,-3.0 can be considered as likely variable,
but up to 9 such sources (on average) might actually be constant. Forty-nine
\cyg sources fall in this category.

\begin{figure}[!ht] \centering \includegraphics[width=8.5cm,angle=0]{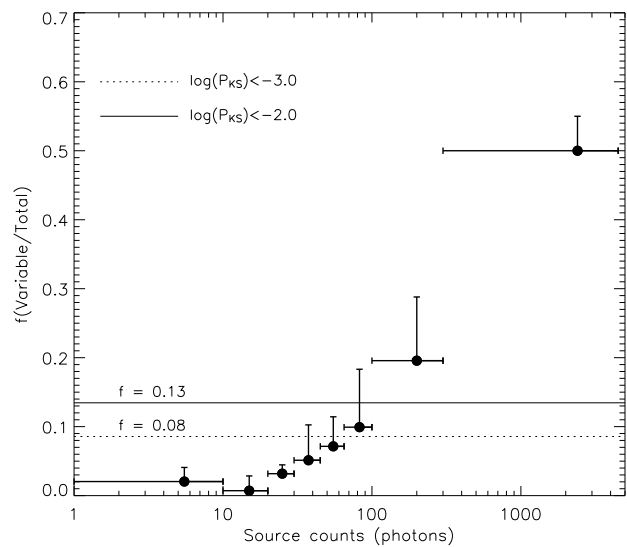}
\caption{Fraction of variable sources as a function of source counts.
Horizontal bars indicate the range of source counts. Filled circles indicate,
for each counts bin, the fraction of sources with log(P$_{\rm KS}$)$<$-3.0 (85
sources in total), while the upper tip of the vertical bars indicate the
fraction with log(P$_{\rm KS}$)$<$-2.0 (134 sources in total).  The continuous
and dashed horizontal lines indicate the overall (average)  variability
fractions obtained for the two confidence levels.} \label{ks} \end{figure}

These numbers of sources in which variability is detected are however a lower
limit to the total number of variable sources in the region for several
reasons: i) most of the observed variability is in the form of flares, i.e.
events that are shorter than our observation and with duty-cycle that may be on
the other hand considerably longer  \citep{2005ApJS..160..423W}; ii) the
sensitivity of statistical tests to time variability of a given relative
amplitude depends critically on photon statistic. This is illustrated in
Figure\,\ref{ks}, where we plot the fraction of variable sources as a function
of source counts: the clear correlation between the two quantities is most
likely due to this statistical bias even though we cannot exclude a real
dependence. 

\begin{figure*}[!ht] \centering \includegraphics[width=18cm,angle=0]{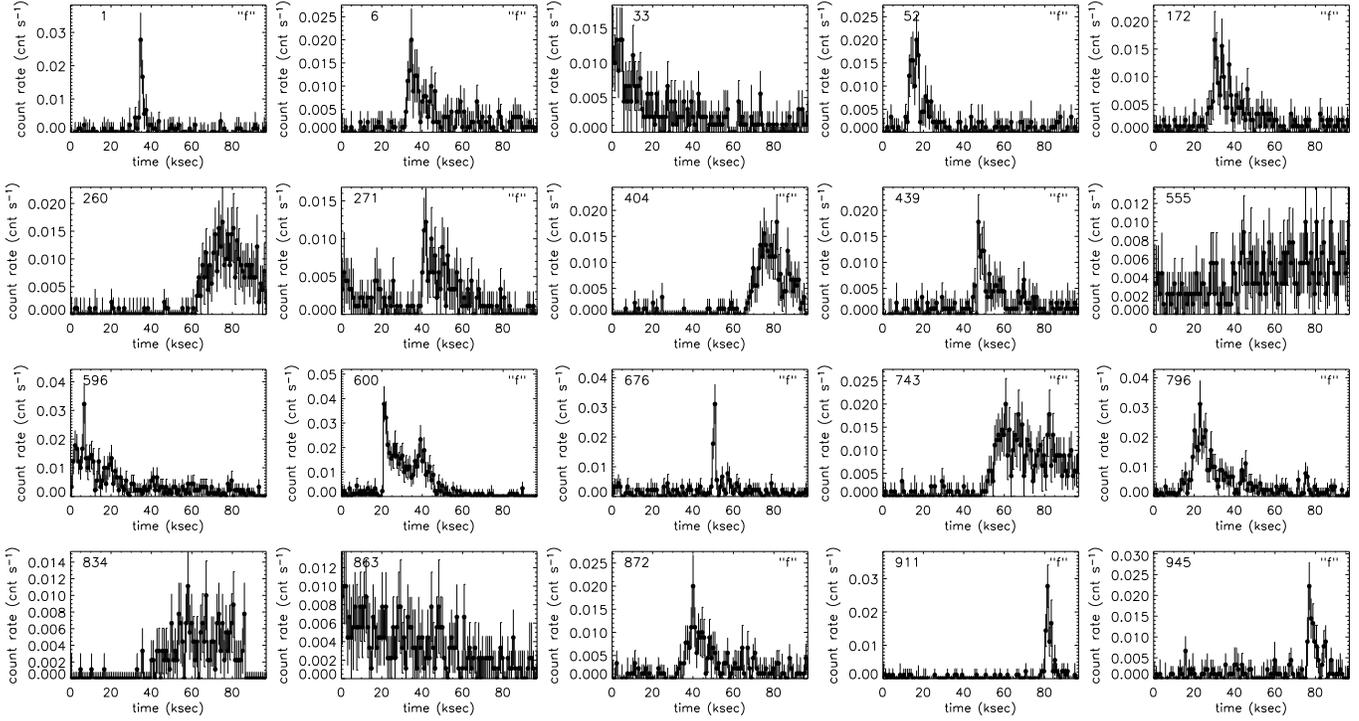}
\caption{Light curves (in the 0.5-8.0 keV band) for 20 sources with variable
emission during our 97.7 ksec \chandra observation. The bin length is 600
seconds. Source numbers are given in the upper-left corner of each panel,
followed by an 'f' for sources classified as flaring. A detailed study of X-ray
properties in flaring sources and the flare frequency in the \cyg region will
be presented in a forthcoming paper (Albacete-Colombo et al. in prep.)}
\label{flr} \end{figure*}

\begin{figure*}[!ht] \centering
\includegraphics[width=8.3cm,angle=0]{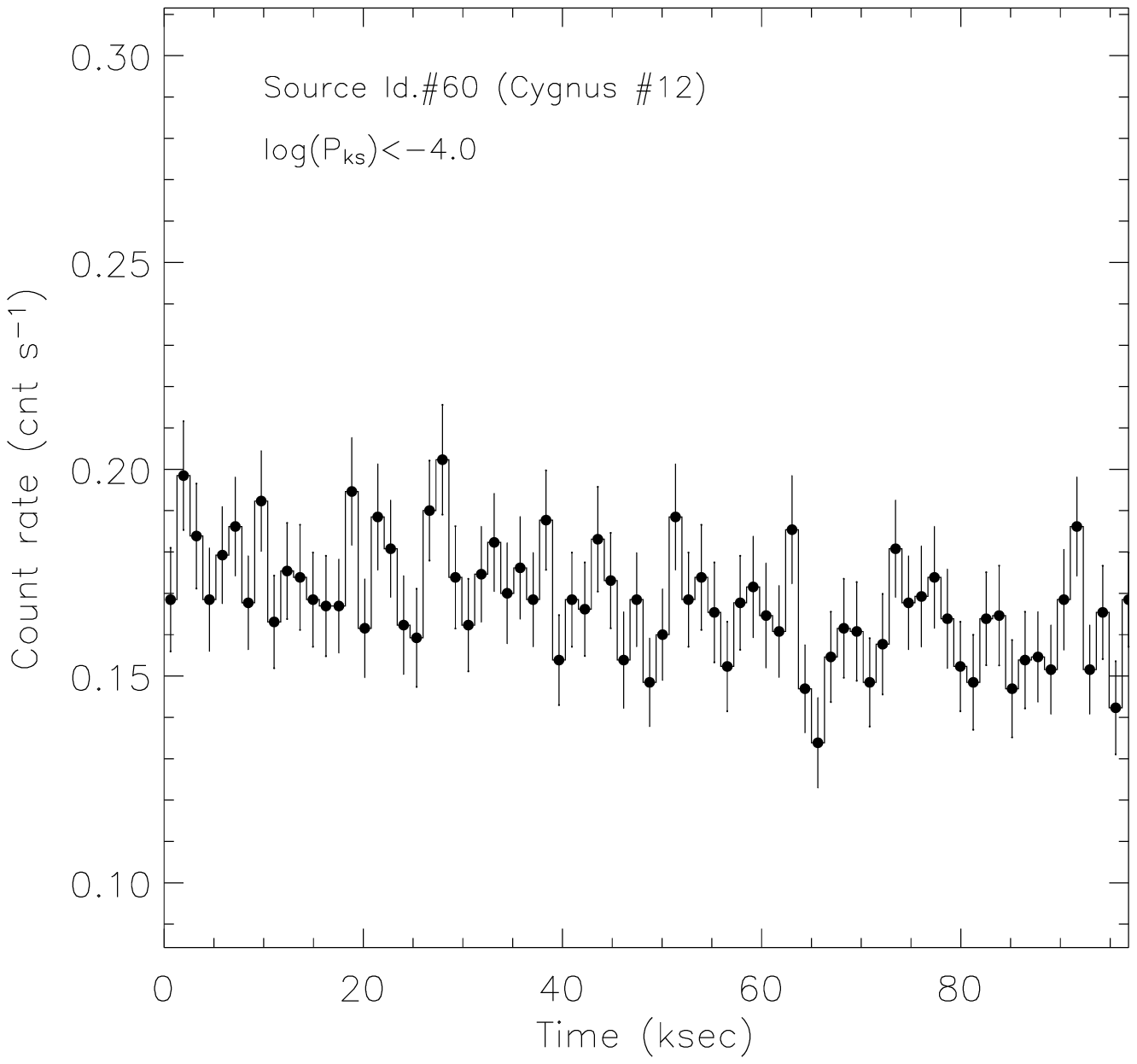}
\includegraphics[width=8.3cm,angle=0]{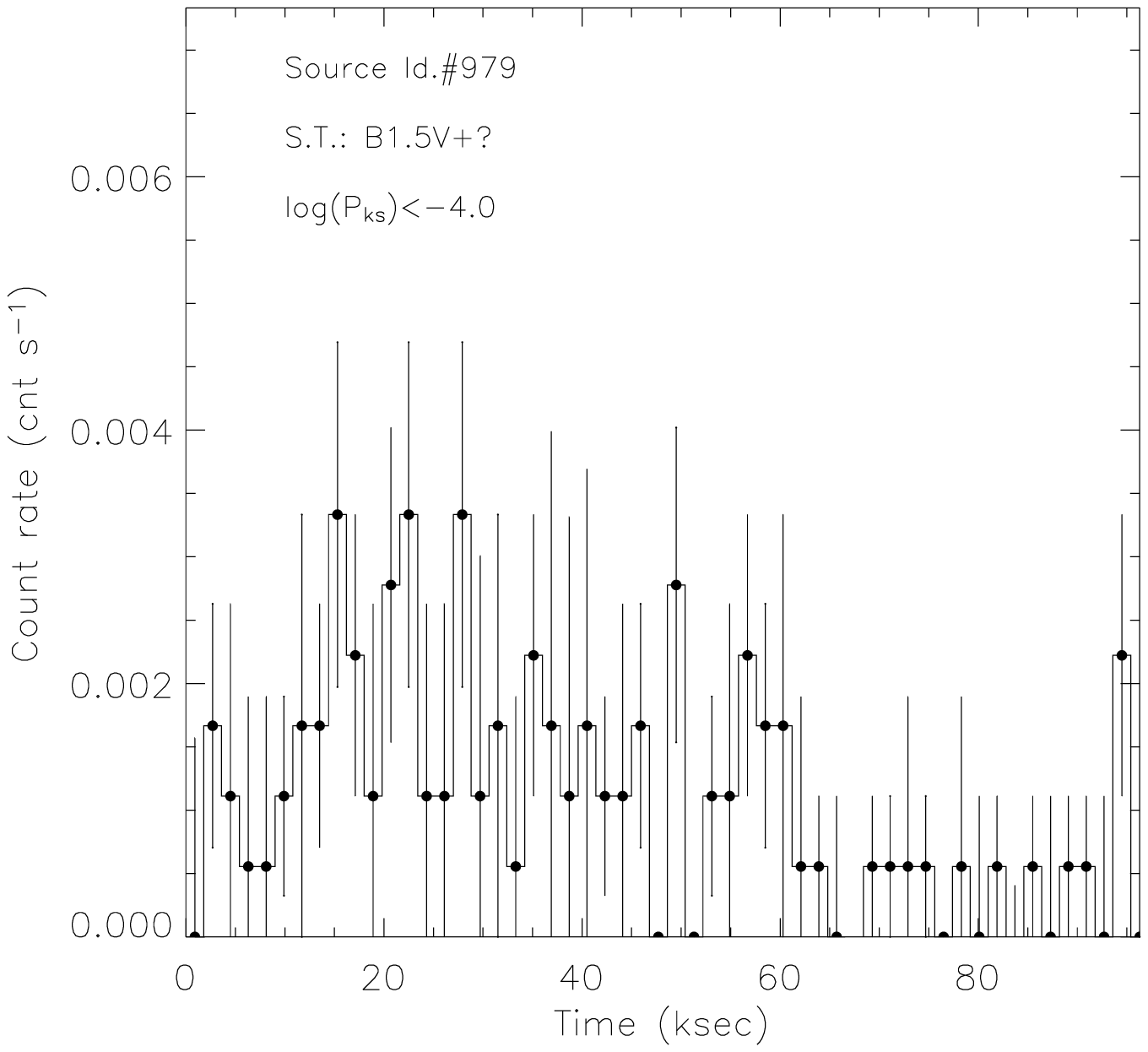}  \caption{Light curves in
the 0.5-8.0 keV band for the two variable early-type stars: {\it Left:}  Source
Id.\,\#60, known as a B5Ie, has been recently classified  a B5$\pm$0.5\,Ia+
star by \cite{2004ARep...48.1005K}. X-ray variability of this star  was unknown
till now. The time bin-size is 900 sec. {\it Right:} Source Id.\,\#979 is
probably a B1.5\,V+'?' binary system \citep{1991AJ....101.1408M}. The time
bin-size here is 1300 sec. Source number and results of the KS test are given in
the legend.}  \label{otime} \end{figure*}

Next, we extracted binned light-curves for each of our \cyg X-ray sources
adopting a bin length of 600 seconds, a compromise between bins that are long
enough to reach a good signal-to-noise ratio per bin for most sources and
sufficiently short to resolve the decay phase of typical flares. Since the
background of our observation is both small (negligible for many sources) and
constant, we did not apply any background subtraction. We inspected the 857
light-curves of source with $>$10 photons, finding flare-like events,
qualitatively defined as a rapid rise and a slow decay, in 98 sources, i.e.
$\approx$ 9.8\% of the cases. These sources are indicated by a $\dag$ in the
last column of Table \ref{AEphot}. Of these 98 sources, 66 ($\sim$\,65\,\%)
have a log(P$_{\rm KS}$)\,$\leq$\,-3.0, while 13 were classified as probably
variable (-2.0\,$<$\,log P$_{\rm KS}$\,$\leq$\,-3.0). The remaining 19 sources
were not detected as variable by the KS test (log P$_{\rm KS}$$>$-2.0) and may
or may not be actually ``flaring''. Figure\,\ref{flr} shows light-curves for 20
variable sources, 14 of which classified as flaring. Some of these sources
(e.g. \#1, \#676 and \#911) experience ``impulsive''  flares with very quick
rises and decay phases of only a a couple of hours.  Others  (e.g. \#33, \#172,
and \#260) show longer (2 to 10 hours) flares. In several instances a second
impulsive event is visible during the exponential decay of a previous flare
(e.g. sources \#6, \#52, \#439, \#600, \#796, and \#945). Other sources (e.g.
\#555, \#834 and \#863) have variable light-curves that bear little resemblance
to typical flares and are instead characterized by slow continuous rises or
decays that might be explained by rotational modulation of non-homogeneously
distributed plasma \citep{2005ApJS..160..450F}.

Finally, we note that, as expected if flares originate from magnetic
reconnection events \citep{2003SSRv..108..577F},  the median photon energies
for flaring sources are generally higher than those of non-flaring sources: the
distribution of median energies for the flaring source peaks at $\overline
E_x$$\approx$2.6 keV, with a 1$\sigma$ dispersion of 0.4 keV, while for
non-variable stars it peaks at $\overline E_x$$\approx$2.1 keV, with a
1$\sigma$ dispersion of 0.3 keV. A similar conclusion can be drawn from the
spectral modeling presented in the next section (\S \ref{sect:specanal}):
flaring sources often require higher temperature models than non-variable ones.

The variability of massive O and early B-type stars is  significantly different
from that of low mass members. Among the 26 OB stars detected in our FOV, 24
are classified as non-variable (log(P$_{\rm KS}$)$>$-2.0), in spite of their
higher than average statistics, having between $\sim$\,40 and 15000 counts,
with a median of $\sim$\,111 counts.  A comparison with Fig. \ref{ks} shows
that the variability fraction of OB stars (7.7\%) is significantly lower than
for the bulk of our sources with similar statistics. This finding agrees with
the common view of X-ray emission from O stars, which is believed to be
unrelated solar-like magnetic activity, and rather explained as the integrated
emission from a large number of small shocks occurring in the strong winds of
these stars \citep{1997A&A...322..878F,1999ApJ...520..833O}. Interestingly
however, two of these sources, \#60 and \#979, are significantly variable, with
log(P$_{\rm KS}$) values lower than -4. Figure \ref{otime} shows their
light-curves.

Source \#60 (Cyg\#12) is a B5\,Ie star \citep{2003ApJ...597..957H}.  The X-ray
light-curve shows a roughly linear decay of the ACIS count rate from
$\approx$0.18 cnt\,s$^{-1}$ to $\approx$0.16 cnt\,s$^{-1}$. The star was
observed by \cite{1998ApJS..118..217W} who did not report variability during
$\sim$125 ksec of non-continuous ROSAT PSPC  observation. This kind of
variability may be similar to the rotational modulation observed on the O7 star
$\theta$ Ori\,C \citep{2005ApJ...634..712G} and there attributed to the
presence of a rigidly rotating magnetosphere (RRM), which may form in magnetic
early-type stars with misaligned magnetic and rotation axis
\citep{2005AIPC..784..239O}.  We defer further discussion about the nature of
the Cyg\,\#12 to \S \ref{sect:reshi}.

Source \#979 is identified with star \#646 in \citet{2003ApJ...597..957H} and
there classified as a B1.5V\,+\,? star, the question mark indicating the
presence of an unresolved faint secondary. The observed variable X-ray emission
is much fainter than in the previous case and might be due to the magnetic
activity of the companion to the B1.5 primary, presumably a lower mass star. 

\section{Spectral analysis}
\label{sect:specanal}

In order to characterize the hot plasma responsible for the X-ray emission of
\cyg stars, and to estimate their intrinsic X-ray luminosities, we analyzed the
ACIS spectra of each of our 1003 sources. Reduced source and background spectra
in the 0.5-8.0 keV band were produced with AE (see \S \ref{sect:extraction}),
along with individual ``redistribution matrices files'' (RMF) and ``ancillary
response files'' (ARF). For model fitting spectra were grouped so to have a
specified number of events in each energy bin. Grouping was tuned to the source
statistics and we chose 2, 5, 7, 10, and 60 counts per channel for sources with
net-counts in the following ranges: [0-40], [40-100], [100-200], [200-500], and
[500-16000].  Spectral fitting of background-subtracted spectra was finally
performed with XSPEC v12.0 \citep{2004HEAD....8.1629A} and our own shell and
TCL scripts to automate the process as described in
\citet{2006astro.ph..4243F}.  Best fit parameters for the chosen models were
found by  $\chi^2$ minimization.

We fit our spectra assuming emission by a thermal plasma, in collisional
ionization equilibrium, as modeled by the {\sc APEC} code
\citep{2001ApJ...556L..91S}. Elemental abundances are not easily constrained
with low-statistic spectra and were fixed at Z=0.3 Z$_\odot$, with solar
abundances taken from \citet{1989GeCoA..53..197A}. The choice of sub-solar
abundances is suggested by several  X-ray  studies of star forming regions
\citep[e.g.][]{2002ApJ...574..258F,2003A&A...401..543P}.  Absorption was
accounted for using the {\sc WABS} model, parametrized by the hydrogen column
density, N$_{\rm H}$ \citep{1983ApJ...270..119M}. 

\begin{figure}[!ht] \centering
\includegraphics[width=9cm,angle=0]{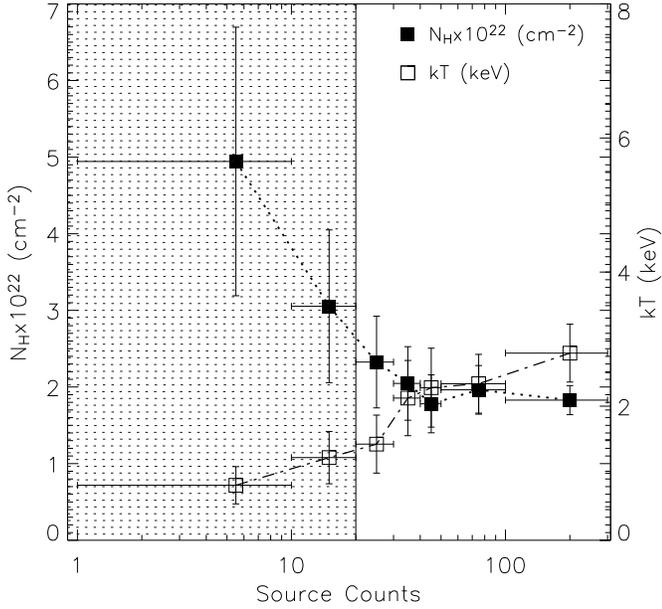} \caption{Median N$_{\rm H}$
and kT vs. source counts for sources fit with one-temperature models. Filled
and empty squares indicate the average values of the $N_H$ and kT in each
considered count range with vertical scales given on the right and left-hand
side, respectively. The extent of the count ranges is indicated by the
horizontal error bars. Vertical error bars indicate the median of the 1$\sigma$
uncertainties  on the two parameters. For this plot we only considered N$_{\rm
H}$  and kT values with formal relative errors smaller than 90\%. The hatched
area below 20 counts indicates the count-range in which spectral fits were
discarded because of strong biases in the best fit parameters. } \label{xcount}
\end{figure}

Except for X-ray sources associated with O- and early B-type stars, which are
discussed separately in \S \ref{sect:reshi}, we fit source spectra with
one-temperature (1T) plasma models using an automated procedure. In order to
reduce the risk of finding a relative minimum in the $\chi^2$ spaces,  our
procedure chooses the best fit among several obtained starting from a grid of
initial values of the model parameters: log(N$_{\rm H}$)\,=\,21.0, 21.7, 22.0,
22.4, 22.7 and 23.0 cm$^{-2}$ and kT\,=\,0.5, 0.75, 1.0, 2.0, 5.0 keV. Best fit
values of log($N_{\rm H}$)\,$<$\,20.8 cm$^{-2}$ were truncated at 20.8 because,
in the 0.5-8.0 keV energy range, ACIS spectra are insensitive to lower column
densities. For the same reason, 154 best fit values of kT turned out to be
$>$10 keV and were truncated at that value.

As noted by many authors 
\citep[e.g.][]{2002ApJ...575..354G,2005ApJS..160..319G,2006astro.ph..4243F},
the spectral fitting of low statistic ACIS sources is problematic  because of a
degeneracy between plasma temperature and absorption. The degeneracy also
results in a systematic bias:  kT values are often underestimated by as much
as  $\sim$50\% while N$_{\rm H}$ values are  overestimated. We investigated
this issue with our data by considering the distributions of the best fit
parameters for source in different count-statistic bins. Figure \ref{xcount}
shows the run of mean kT and N$_{\rm H}$ with source counts for spectra fitted
with 1T models. The  systematic decrement of N$_{\rm H}$ and the increment of
kT with increasing source statistics are hardly explainable as physical effects
and rather indicate that the spectral parameters obtained for sources with less
than $\sim$20 photons are ill-constrained.

The uncertainty on the N$_{\rm H}$ is particularly serious as it implies large
and systematic uncertainties on the absorption-corrected X-ray luminosities. In
order to reduce the risk of erroneous results, we discarded from the following
analysis results for source with $<$20 net photons, a total of 423 sources. 
Table \ref{xspec} presents the results of the automated 1T spectral fits for
554 sources (the results of the spectral fitting for the 26 sources associated
with know OB stars will be presented in  \S \ref{sect:reshi}, Table \ref{OB}).
We list source numbers (column 1) from Tables 1 and 2, background subtracted
counts in the spectra (2), the best fit hydrogen column densities and their
1\,$\sigma$ errors (3), the plasma temperatures and their 1\,$\sigma$ errors
(4), and the emission measures (5). Column (6) and (7) give the reduced
$\chi^2_\nu$ for the spectral fits and the relative degrees of freedom,
respectively.  

\begin{table}
\small
\begin{center}
\caption{Results of X-ray spectral fits: first rows. The complete version 
is available in the electronic edition of A\&A.}
\label{xspec}
\begin{tabular}{lllllll} 
\multicolumn{7}{l}%
{{\bfseries}}\\
\hline \hline
\multicolumn{1}{l}{N$_{\rm x}$} &
\multicolumn{1}{l}{Cnts} &
\multicolumn{1}{l}{log(N$_{\rm H}$)} &
\multicolumn{1}{l}{kT} &
\multicolumn{1}{l}{log(EM)} &
\multicolumn{1}{l}{{Stat.}} &
\multicolumn{1}{l}{{dof}} \\
\multicolumn{1}{l}{\#} &
\multicolumn{1}{l}{(ph.)} &
\multicolumn{1}{l}{(cm$^{-2}$)}&
\multicolumn{1}{l}{(keV)}&
\multicolumn{1}{l}{(cm$^{-3}$)} &
\multicolumn{1}{l}{($\chi^2_\nu$)} &
\multicolumn{1}{l}{} \\
\hline
   1  &      67  &  22.13$\pm$0.26  &  9.99           &  53.70  &   0.34  &	17  \\
   2  &      72  &  21.05$\pm$0.34  &  0.74$\pm$0.12  &  53.40  &   1.54  &	17   \\
   3  &      35  &  21.75$\pm$0.03  &  9.99           &  53.40  &   0.38  &	 9   \\
   5  &      55  &  22.15$\pm$0.27  &  6.05$\pm$8.52  &  53.40  &   0.81  &	14   \\
   6  &     216  &  22.13$\pm$0.68  &  5.52$\pm$3.06  &  54.10  &   0.96  &	19  \\
   7  &      49  &  21.86$\pm$0.04  &  9.99           &  53.40  &   0.78  &	15   \\
   8  &      33  &  21.70$\pm$0.08  &  9.99           &  53.40  &   0.51  &	 9   \\
   9  &      64  &  21.44$\pm$0.03  &  9.99           &  53.40  &   0.76  &	14   \\
  10  &      50  &  21.60$\pm$0.10  &  1.04$\pm$0.23  &  53.40  &   1.00  &	12   \\
  12  &      99  &  22.14$\pm$0.47  &  2.19$\pm$0.91  &  53.88  &   1.02  &	24   \\
  13  &      44  &  22.42$\pm$0.29  &  3.74$\pm$3.87  &  53.70  &   1.20  &	10   \\
  14  &      75  &  22.51$\pm$0.30  &  9.99           &  53.70  &   1.10  &	19   \\
  15  &      60  &  21.90$\pm$0.24  &  6.85$\pm$8.19  &  53.40  &   0.66  &	14   \\
  16  &      30  &  22.53$\pm$0.10  &  2.72$\pm$2.67  &  53.70  &   0.82  &	 7   \\
  17  &    3173  &  22.16$\pm$0.25  &  1.34$\pm$0.64  &  55.65  &   1.57  &    252   \\
  19  &      25  &  22.48$\pm$0.15  &  3.43$\pm$4.19  &  53.40  &   0.56  &	 6   \\
  20  &      49  &  22.26$\pm$0.30  &  3.04$\pm$1.86  &  53.70  &   0.72  &	11   \\
\hline
\end{tabular}
\smallskip
\end{center}
\end{table}

\begin{figure*}[ht] \includegraphics[width=8.6cm,angle=0]{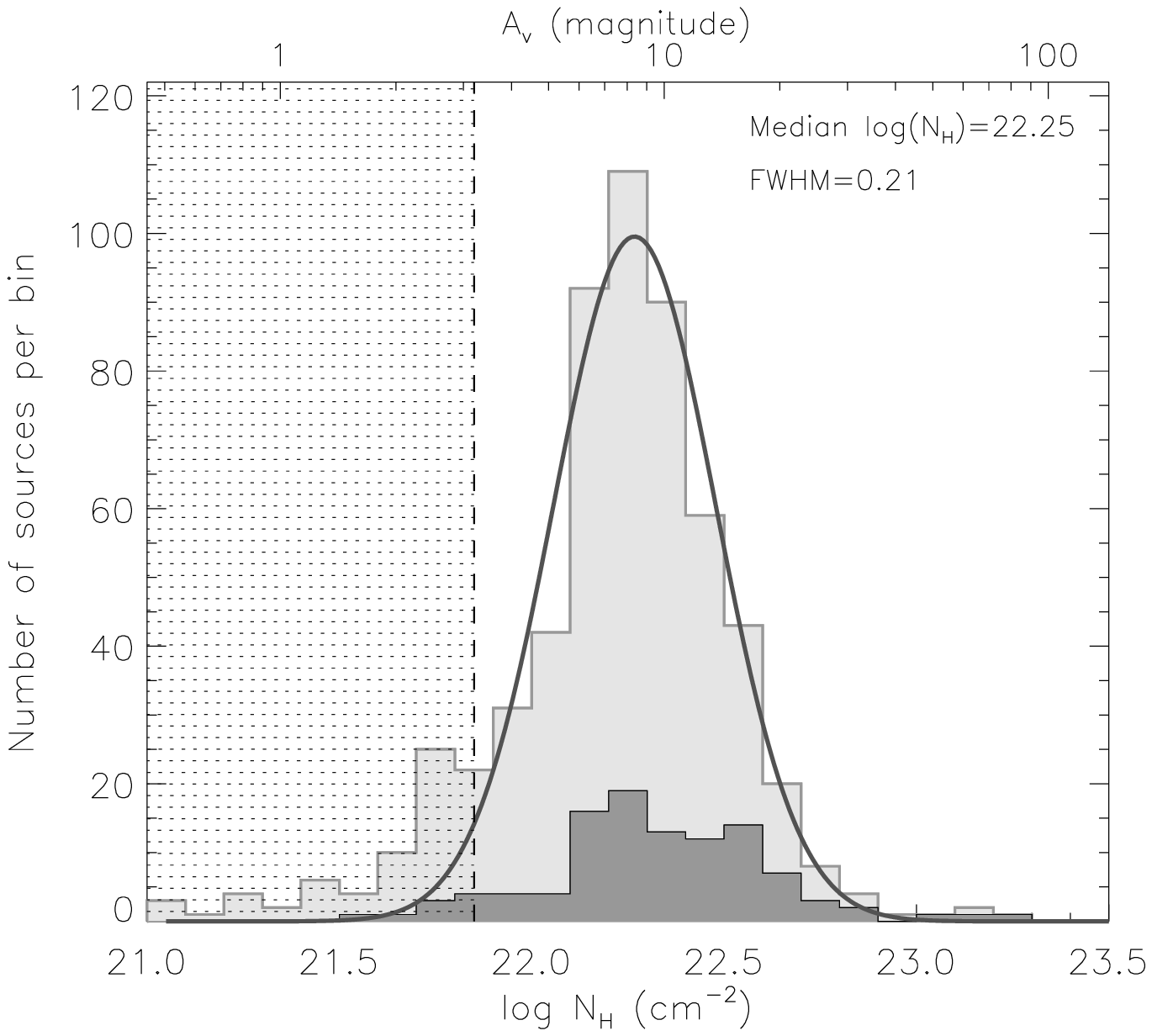}
\includegraphics[width=8.6cm,angle=0]{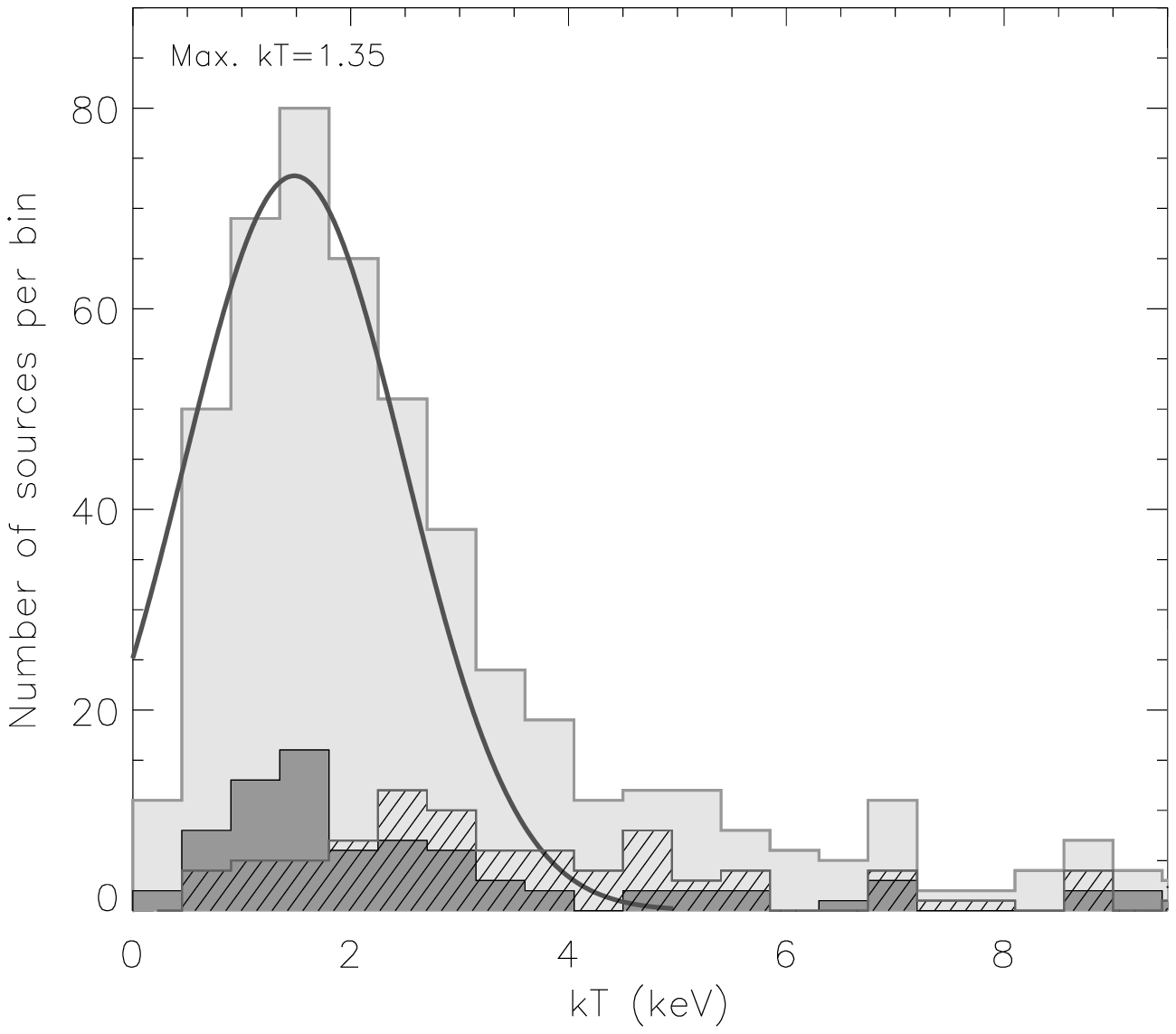} \caption{(a):
Distributions of absorbing columns, N$_{\rm H}$, with corresponding visual
absorption scale at the top. The light gray area refers to sources with
net-counts greater than 20. The maximum of the distribution is at log N$_{\rm
H}$\,$\sim$\,22.25 cm$^{-2}$ with a Full Width Half Maximum (FWHM) of about
0.21\, dex. The shaded area on the left indicates an excess of low-absorption
sources with respect to the log-normal distribution that best describes the
histogram (the continuous curve), possibly associated with foreground stars.
The darker gray histogram refers to the subsample of sources without a 2MASS
counterpart. (b): same as panel (a) for plasma temperatures (kT) with, in
addition, the distribution for ``flaring'' sources (hatched histogram). The
peak of the overall distribution is at $\sim$1.35 keV. } \label{par_xspec} 
\end{figure*}

The hydrogen column densities derived from X-ray spectral fitting depend of the
interstellar material in the line of sight to the \cyg cloud and on the
location of stars within the cloud material of \cyg. Figure \ref{par_xspec}(a)
shows the distribution of log N$_{\rm H}$ values for the 580 sources with more
than 20 counts. They appear to be normally distributed with a median
log\,$N_{\rm H}$\,$\sim$\,22.25 (cm$^{-2}$) and a FWHM of $\sim$ 0.21 dex.  The
shaded area in Figure \ref{par_xspec}(a) indicates an apparent excess of
relatively unabsorbed sources (log N$_{\rm H}<$21.8 cm$^{-2}$) with respect to
the log-normal distribution of the bulk of the sources. These $\sim$23 sources
are likely associated with foreground stars. In fact we noticed that most of
these sources seem to be spatially distributed uniformly in the FOV of the \cyg
region. 

Another interesting observation is that the N$_{\rm H}$ distribution of the 106
X-ray sources without near-IR counterparts  (dark gray histogram in
Fig.\,\ref{par_xspec}) seems to be skewed toward higher values with respect to
global distribution. For example, while only 11\% of the X-ray sources with
counterparts have N$_{\rm H}>22.5$, this is true for 29 (27\%) of the
unidentified sources. 

According to the relationship between N$_{\rm H}$ and A$_{\rm V}$,  N$_{\rm
H}$=2.2$\times$10$^{21}$\,\Av cm$^{-2}$  \citep{1996Ap&SS.236..285R}, the
median N$_{\rm H}$ of all detected sources with more than 20 counts corresponds
to \Av $\sim$\,8.1$^{12.8}_{5.1}$ mag. If we only consider the sources with
near-IR counterparts (and $>$ than 20 counts) we obtain log N$_{\rm H}$=22.23,
which corresponds to \Av=7.7 mag., i.e. in reasonable agreement with the  value
calculated from the NIR CMD (\Av$\approx$7.0 mag). 

Figure \ref{par_xspec}(b) shows the distribution of plasma temperatures: it
peaks at $\sim$1.35 keV, has median $\sim$2.4 keV and shows an extended hard
tail. Variable sources (log\,P$_{\rm KS}<-2$) have harder spectra with median
kT=3.3\,keV and flaring sources (see hatched histogram in Fig.
\ref{par_xspec}b) are even harder with median kT=3.75 keV.

Finally we derived unabsorbed X-ray luminosities for each of our sources. For O
and B-type stars, discussed in detail below (\S \ref{sect:reshi}) luminosities
were derived from individual spectral fits. For the other sources, given the
considerable uncertainties in the absorption estimates, especially at the faint
end, we preferred not to use the results of spectral fits individually. Rather,
we computed a single count-rate to L$_{\rm X}$ conversion factor that takes
into consideration the average intrinsic source spectrum and interstellar
absorption. This conversion factor, computed as the median ratio between the
individual unabsorbed L$_{\rm X}$'s (from the best fit spectral models) and the
source count-rates is 7.92 $10^{33}$ ergs.

\section{Results}
\label{sect:results}

We now discuss the implications of our data for the understanding of X-ray
emission from young stars and for the study of the \cyg stellar population. We
split the discussion according to stellar mass and, correspondingly, expected
X-ray emission mechanism. We first discuss known O and early B ($<$B2) stars,
i.e. massive stars ($\rm M\gtrsim 12M_\odot$) expected to generate X-rays in
their powerful stellar winds. Next, using our mass estimates based on the
J-band magnitude (\S \ref{sect:NIRprop}), we define and study the properties of
two subsamples of intermediate and low-mass X-ray detected stars. We base the
distinction between the two groups on the model-predicted thickness of the
convective layer, whose presence has been often found to be correlated with
solar-like coronal activity. Stars with small or absent convective layers are
instead believed to be unable to sustain coronal activity, although the matter
is somewhat controversial. Adopting the 2Myr isochrone from the SDF models, we
observe that the thickness of the convective layer relative to the stellar
radius, R$_{\rm c}$, decreases rapidly, from $\sim$5\% to $<$1\%, as the
stellar mass increases from 3 to 3.3M$_\odot$. This rather sharp boundary
however lies right in the 2.8-5.3M$_\odot$ range in which our mass estimates
are degenerate and therefore unconstrained (see \S \ref{sect:NIRprop}).  In
order to resolve this ambiguity, we define the intermediate-mass  range as
2.8-10 M$_\odot$, resulting in the selection of 57 stars\footnote{There are
only 28 stars with masses in the 2.8-5.3M$_\odot$ range and for which this
interval formally corresponds to the uncertainty on the mass estimate. We
therefore expect a small fraction of the 57 intermediate mass stars to lie in
the 2.8-3.0M$_\odot$ range and to have a substantial model-predicted convective
envelope.}. Low mass stars are finally defined as stars with M$<$2.8M$_\odot$
and are predicted to have substantial convective envelopes (R$_c>$18\%). 578
counterparts to X-ray sources fall in this category. Because of the sensitivity
limit of 2MASS only 25(4) sources of these stars have masses below
0.5(0.4)M$_\odot$, which we may consider as our approximate detection limit.
Our {\em completeness} limit is however probably higher, at $\sim$1M$_\odot$,
corresponding J=15.8, i.e. roughly the completeness limit of 2MASS. An
histogram of the logarithm of stellar masses also peaks at $\sim$1M$_\odot$,
confirming the above estimate given that the mass function usually increases
down to lower masses.

\begin{figure}[!ht] \centering
\includegraphics[width=8.8cm,angle=0]{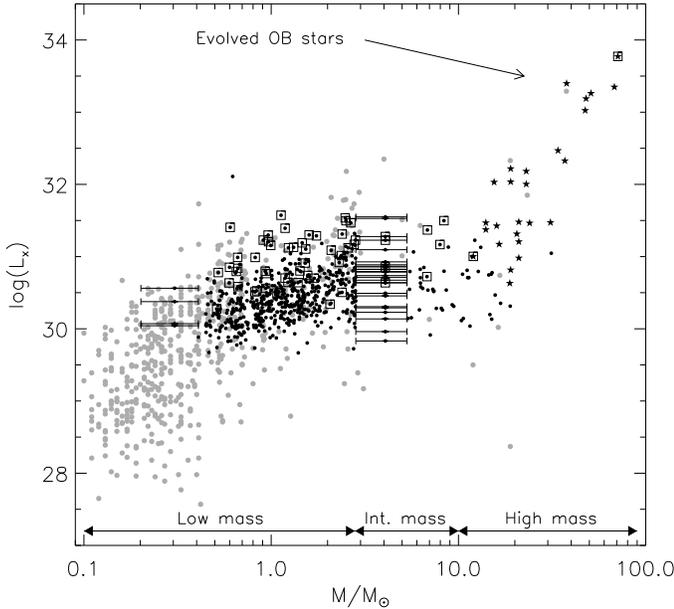} \caption{Log M/M$_\odot$
vs. log L$_{\rm x}$ for \cyg sources with NIR counterparts (black circles and
star-symbols) and for ONC sources observed by the COUP project (gray circles).
Variable sources are indicated by empty squares, known O- and early B-type
stars by star-symbols.} \label{lx-mass} \end{figure}

Figure \ref{lx-mass} shows the L$_X$-mass scatter plot for 683 X-ray sources in
all three mass ranges. We distinguish with different symbols stars of low- and
intermediate- mass, as determined from the J-band magnitude, and massive stars
with known O- and early B- spectral types\footnote{A number of X-ray detected
stars with no spectral type appear to be of high mass but with lower X-ray
luminosities with respect to those with spectral types. In the discussion of
high mass stars, however, we will only consider stars with spectral types.}.
Note that stars in the 2.8-5.3M$_\odot$ and 0.2-0.4\,M$_\odot$ ranges are
plotted with horizontal error bars spanning the entire ranges, because the
J-band vs. mass relation was there found to be degenerate. In the plot we also
indicate the position of 57 time-variable X-ray sources. They appear to be
brighter than other sources of the same mass, probably because variability is
more readily detected in sources with high photon statistics (see \S
\ref{sect:var}).

\subsection{High-mass stars}
\label{sect:reshi}

The \cyg region is one of the most massive SFR in the Galaxy, with
$\sim$2600$\pm$400  OB star members \citep[][]{2000A&A...360..539K}. The
presence of a variety of evolved stars suggests that the star formation process
was non-coeval \citep[][]{1991AJ....101.1408M,2003ApJ...597..957H}. In fact,
out of a total of 20 O-type and 13 B-type stars lying in our \cyg ACIS FOV, 7
and 3, respectively, are evolved (luminosity class: I-III). In this respect it
is interesting to note that, while we detect X-ray emission from all O stars
regardless of evolutionary status, we have only detected 6 B stars, among which
all the evolved ones. The detection fraction of evolved B star is thus 100\%
while it is only 33\% (3 our of 9\footnote{One undetected B stars has unkown
luminosity class. }) for un-evolved ones.

The most widely accepted explanation for the X-ray emission of single O and
early B-type stars invokes multiple small-scale shocks in the inner layers of
their radiation-driven stellar winds \citep[e.g.][]{1997A&A...322..878F}.
Recent theoretical and observational results however support, for  several OB
stars, another physical plasma heating model involving strong magnetic fields: 
the magnetically channeled wind shock (MCWS) model
\citep[][]{2003ApJ...595..365S,2005AIPC..784..239O,2005ApJ...634..712G}.
Moreover, binary systems in which both components are O- and early B-type
stars, can produce intense thermal X-ray emission from wind-wind interactions
as well as non-thermal X-ray emission from Inverse Compton scattering. 

We modeled the spectra of O and B stars with one- or two-temperature thermal
plasma emission models ({\sc apec}), absorbed by neutral interstellar and
circumstellar material ({\sc wabs}). Because the dense stellar winds of evolved
OB stars are affected by the strong stellar UV/EUV ionizing radiation, the
additional absorption of X-rays by a partially ionized stellar wind should be
also considered.   For evolved stars we then decided to model the combined
effect of ISM plus wind material with a {\it warm} absorption model \citep[{\sc
wabs\,$\times$\,absori}, see e.g.][]{1998ApJS..118..217W}. For sources \#60,
\#544, and \#729, we found that the inclusion of a warm absorber in the
spectral model yields satisfactory fits of the soft part of the spectra
($\le$1.2 keV), reducing the $\chi^2_\nu$ from 2.1 to 1.4, from 2.7 to 1.3, and
from 2.2 to 1.2, for the three sources respectively. The spectra of other
evolved sources did not instead support the presence of the warm absorber,
although in many cases the spectra have too low statistics to be conclusive.
Metal abundances were fixed at Z\,=\,0.3\,Z$_\odot$  in fitting faint sources
($<$\,100 ph.) and  left as a free parameter for brighter sources. When the
best fit abundance was lower than Z\,=\,0.1\,Z$_\odot$ we repeated the fit with
Z fixed at this value.

Figure \ref{obspectra} shows three notable examples of spectral modeling and
the best-fit parameter values for the 26 detected OB stars are presented in
Table \ref{OB}. We list: X-ray source numbers from Table 1 (column 1),
identification numbers and spectral types from \citet{2003ApJ...597..957H} (2
and 3), reduced $\chi^2$ of the best fit (4), fraction of pile-up as reported
by AE (as discussed \S \ref{sect:extraction}, if f(Pile-up)$>$2\% we used an
annular photon extraction region that excludes the peak of the PSF), the N$_H$
of the {\it cold} and {\it hot} absorption components from {\sc wabs} and {\sc
absori}, respectively (6 and 7), plasma temperatures of the two components (8
and 9), metal abundances (10). Column (11) finally gives the un-absorbed X-ray
flux in the 0.5\,-\,8.0 keV energy band.

\begin{table*}
\begin{center}
\caption{X-ray spectral parameters of the \cyg OB-type stars}
\begin{tabular}{l ll l l ll l l l l l}
\hline \hline
Src. &\multicolumn{2}{l}{Optical Parameters} &$\chi_\nu^2$&Pile-up&\multicolumn{2}{l}{N$_{\rm H}$[$\times$10$^{22}$] [cm$^{-2}$]}&kT$_1$ & kT$_2$ & Abundance   &  Flux [10$^{-13}$]\\
\cline{2-3} \cline{6-7}
\#   &Hanson\#&S.T.&&(\%)&{\sc wabs}&{\sc absori}&[keV]&[keV]&[Z$_\odot$]&[erg\,s$^{-1}$\,cm$^{-2}$]\\
\hline
60$^b$  &304&B5\,Ie     	&1.4&$\sim$18	&2.15$\pm$0.15	&0.87$\pm$0.12	&0.63$\pm$0.03&1.56$\pm$0.11&0.43$\pm$0.04&235.4$^{255.4}_{201.7}$\\
133     &339&O8.5\,V    	&1.1&no		&0.24$\pm$0.18	&$---$		&1.53$\pm$0.68&$---$        &(0.3)   	  &0.11$^{0.19}_{0.07}$\\
253     &378&B0\,V      	&1.0&no		&3.57$\pm$0.09	&$---$		&0.43$\pm$0.15&$---$        &(0.3)        &2.00$^{2.21}_{1.42}$\\
315     &390&O8\,V      	&0.8&no		&3.30$\pm$0.08	&$---$		&0.44$\pm$0.10&$---$        &0.15$\pm$0.09&0.64$^{0.73}_{0.51}$\\
433     &417&O4\,III(f)    	&1.5&$\lesssim$2&2.30$\pm$0.07	&$---$		&0.39$\pm$0.06&1.94$\pm$0.73&0.60$\pm$0.34&42.1$^{45.3}_{34.6}$\\
456     & 421 & O9.5\,V 	&1.4&no		&1.45$\pm$0.08	&$---$ 		&1.71$\pm$1.02&$---$ 	    &(0.30) 	  & 1.06$^{0.88}_{1.09}$\\ 
488$^b$ &431&O5\,If     	&1.3&$\sim$12	&2.09$\pm$0.07	&$---$		&0.79$\pm$0.05&2.82$\pm$0.61&0.65$\pm$0.10&72.6$^{75.5}_{67.5}$\\
530     &448&O6\,V      	&1.4&no  	&1.97$\pm$0.34	&$---$		&0.65$\pm$0.20&$---$        &(0.1)        &1.18$^{1.39}_{0.96}$\\
537     &455&O8\,V      	&1.1&no  	&2.19$\pm$0.07	&$---$		&0.61$\pm$0.17&$---$        &(0.1)        &1.20$^{1.28}_{1.07}$\\
544     &457&O3\,If     	&1.3&$\lesssim$2&2.12$\pm$0.06	&0.56$\pm$0.09	&0.39$\pm$0.14&1.32$\pm$0.19&0.97$\pm$0.3 &88.8$^{98.6}_{77.5}$\\
562     &462&O6.5\,II   	&1.0&no  	&2.52$\pm$0.48	&$---$		&0.40$\pm$0.12&$---$        &0.14$\pm$0.1 &8.45$^{9.1}_{6.4}$\\
568$^a$ &465&O6If\,+\,O5.5III(f)&1.4&$\sim$15	&2.94$\pm$1.25	&$---$		&0.91$\pm$0.05&2.64$\pm$0.07&0.52$\pm$0.07&61.3$^{64.9}_{57.1}$\\
584     &470&O9.5\,V    	&0.8&no  	&1.85$\pm$0.07	&$---$		&0.44$\pm$0.18&$---$        &(0.3)  	  &3.49$^{3.98}_{2.89}$\\
597     &473&O8.5\,V    	&1.0&no  	&1.88$\pm$0.09	&$---$		&0.30$\pm$0.22&$---$        &1.07         &3.00$^{3.78}_{2.38}$\\
615     &480&O7.5\,V    	&1.0&no  	&2.73$\pm$0.06	&$---$		&0.50$\pm$0.17&$---$        &(0.1)        &4.01$^{4.43}_{3.42}$\\
625     &483&O5\,If     	&1.0&no  	&1.00$\pm$0.08	&$---$		&0.55$\pm$0.06&$---$	    &0.60$\pm$0.21&99.4$^{107.3}_{86.4}$\\
626     &485&O8\,V     	 	&1.1&no  	&1.65$\pm$0.04	&$---$		&0.88$\pm$0.23&$---$        &(0.1)        &0.38$^{0.53}_{0.21}$\\
686     &507&O8.5\,V    	&1.1&no  	&2.39$\pm$0.08	&$---$		&0.35$\pm$0.11&$---$        &(0.3)        &4.16$^{5.04}_{3.28}$\\
729$^a$ &516&O5.5\,V(f)    	&1.1&$\lesssim$4&0.81$\pm$0.05	&0.51$\pm$0.32	&1.98$\pm$0.12&$---$	    &0.38$\pm$0.09&11.7$^{12.4}_{10.1}$\\
779     &534&O7.5\,V  		&1.2&no  	&1.16$\pm$0.03	&$---$		&2.28$\pm$0.55&$---$	    &(0.1)	  &6.05$^{7.74}_{4.91}$\\
817     &556&B1\,Ib     	&0.9&no  	&1.57$\pm$0.08	&$---$		&0.18$\pm$0.07&$---$        &(0.3) 	  &1.20$^{1.53}_{0.94}$\\
881     &588&B0\,V      	&0.9&no  	&1.84$\pm$0.05	&$---$		&0.42$\pm$0.20&$---$        &(0.3)        &0.81$^{1.02}_{0.71}$\\
895     &601&O9.5\,III 		&1.0&no  	&1.98$\pm$0.07	&$---$		&0.38$\pm$0.21&$---$        &(0.3)        &0.72$^{0.92}_{0.50}$\\
971     &642&B1\,III    	&0.9&no  	&1.73$\pm$0.05	&$---$		&0.96$\pm$0.45&$---$        &(0.3)        &0.14$^{0.30}_{0.09}$\\
979     &646&B1.5\,V\,+?	&1.5&no  	&2.90$\pm$1.24  &$---$		&1.92$\pm$1.01&$---$	    &(0.3)        &0.22$^{0.63}_{0.29}$\\
1001    &696&O9.5\,V    	&0.8&no  	&0.71$\pm$0.02	&$---$		&0.54$\pm$0.17&$---$        &(0.3)        &0.71$^{0.83}_{0.42}$\\
\hline
\end{tabular}
\label{OB}
\end{center}
\smallskip

Notes: (1) Abundance values in parenthesis indicate cases for which abundances
were fixed in the spectral fit (see text), (2) Absorption-corrected fluxes are
calculated in the 0.5-8.0 keV energy range (3) Three bright sources,  \#60,
\#488, and \#568, suffer of pile-up, and we extracted the spectra from annular
rings (see \S \ref{sect:extraction}) that exclude the central parts of the
PSFs. ($^a$) the 6.4+6.7 keV FeK$_\alpha$ complex is observed; ($^b$) the
Fe{\sc i} 6.4 keV line is not detected, but the 6.7 keV seems to be present.

\end{table*}

The median absorption of the 26 OB stars is log $N_{\rm H}$=22.19 cm$^{-2}$,
very similar but slightly lower than the median value of lower mass stars
(22.23, see \S \ref{sect:lm_stars}). It is  tempting to attribute the lower
absorption of O stars to the sweeping of the cloud material by strong winds. A
two-sided KS test comparing the distribution of $N_{\rm H}$ for O and B stars
with that of lower mass ones gives, however, a null result.

Most of the O- and early B- type stars are well fit by a soft thermal emission
model with kT ranging between 0.5 and 0.7 keV ($\sim$5.8-8.1 MK), in agreement
with the predictions of the wind shock model
\citep{1980ApJ...241..300L,1988ApJ...335..914O}. The spectra of some sources,
however, require a second (hard) thermal component, which is not predicted by
the wind shock model. The presence of this component seems to depend on the
stellar evolutionary status: out of the 10 evolved O- and early B-type stars
(spectral types B5\,Ie, O4\,III, O5\,If, O3\,If, O6\,If\,+O5.5III(f)) 5 (50\%)
show evidence of a hard thermal emission with kT roughly between 1.3 and 3.0
keV (14-35 MK), while among the 16 class V stars, the hot component is only
required in 3 cases (i.e. 23\%). Several physical mechanisms have been proposed
to explain the hard emission  observed in some of our OB stars:\\
\begin{itemize}  \item {\it Magnetically Confined Wind Shocks} (MCWS). X-ray
emission from MCWS is predicted by simulations of radiation-driven winds in the
presence of a magnetic field \citep{2002ApJ...576..413U} and it has most likely
been observed in O-type stars  \cite[e.g. $\theta^1$Orionis\,C,
][]{2005ApJ...634..712G}. The model predicts that the overall degree of plasma
confinement is determined by a single dimensionless parameter $\eta_\star
\equiv {B_{\rm eq}^2 R_\star \over \dot M \,v_\infty}$, where B$_{\rm eq}$
is the equatorial dipole  magnetic intensity at the stellar surface, R$_\star$
is the stellar  radius, $\dot M$ is the mass-loss rate, and v$_\infty$ is the
terminal wind speed. Simulations with typical parameters for OB stars predict
high post-shock temperatures of about 20\,-\,30 MK (1.7\,-\,2.6 keV).
Furthermore, in the case of oblique rotators, i.e. when the rotational and
magnetic axes are misaligned, rotational  modulation of the X-ray emission can
be observed. This might be the case of our source B5\,Ie (Id.\#60 $\equiv$
Cyg\#12) for which kT$_2$=1.56 keV and the light-curve shows, during our
observation, a gradual decay (c.f. Figs. \ref{otime} and \ref{obspectra}).

\item {\it Highly compressed wind shocks} in evolved O-type stars or {\it
colliding winds} in the case of early-type binary  systems
\citep{1992ApJ...386..265S,2003AdSpR..32.1161R}.   Our source \#568 (Cyg\#8) is
an O6If\,+\,O5.5III(f) supergiants-giant pair, whose X-ray spectrum indicates
hot thermal plasma at kT$_2$\,=\,2.64$\pm$0.07 keV. A colliding wind scenario
is moreover suggested by the presence of the  FeK$_\alpha$ complex (6.4+6.7
keV) in its spectrum\footnote{The temperatures reached by hydrodynamic shocks
in the winds of single stars are usually not high enough to produce significant
Fe K$_\alpha$ emission. In wide early-type binary systems, however, stellar
winds collide with velocities close to $v_{\infty}$ and are thus heated to
sufficiently high temperatures. In such cases, the cooler surrounding wind
material, excited by the high energy radiation, can produce the 6.4 keV
fluorescent Fe line. The observation of the Fe-complex can thus be a used as
diagnostic for colliding-wind binaries \citep{2003AdSpR..32.1161R}.} (c.f. 
Fig. \ref{obspectra}). The other two O-type super-giants, sources \#544
(O3\,If) and \#625 (O5\,If, see Fig. \ref{obspectra}), although requiring a
second thermal component, show rather soft spectra and no FeK$_\alpha$ (6.4+6.7
keV) complex.

\item {\it Inverse Compton scattering} of photospheric UV photons by
relativistic particles that are Fermi-accelerated in shocks within the chaotic
stellar winds of OB supergiants \citep[c.f.][]{1991ApJ...366..512C}. The actual
production of the needed relativistic electrons by stellar wind shocks is,
however, still in doubt. We can likely rule out this possibility for our \cyg
OB stars, given that all of them are well fit by thermal emission models in
which the lines of several elements give a significant contribution.

\end{itemize}

\begin{figure}[!h] \centering
\includegraphics[width=9cm,angle=0]{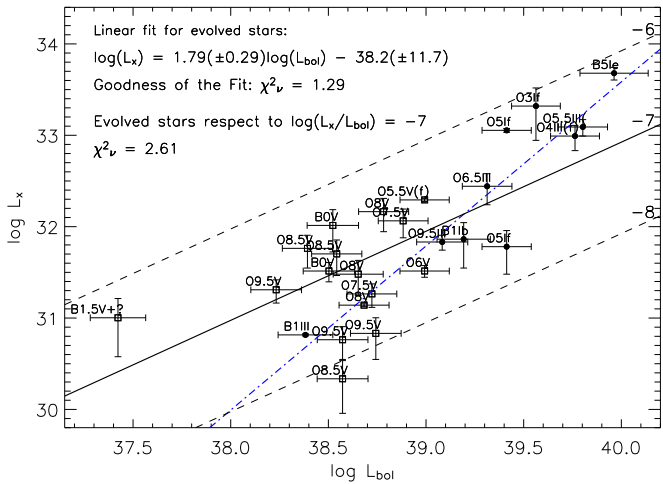} \caption{X-ray versus
bolometric luminosity for main-sequence and evolved OB-type stars (open
and filled symbols, respectively) in the \cyg\, region. The dotted and
continuous lines indicate log L$_{\rm x}$/L$_{\rm bol}$= -6, -7, and -8.
Evolved stars (i.e. giants and super-giants) seems to follow a power-law
relation with index $\sim$1.8 (dashed line).} \label{lxlbol} \end{figure}

One of the most intriguing and so far un-explained observational results
regarding the X-ray emission of O- and early B- type stars is the existence of
a simple proportionality between the X-ray luminosity  (L$_{\rm x}$), and the
bolometric luminosity (L$_{\rm bol}$) \citep{1997A&A...322..167B}. We explore
this relationship for \cyg\, O- and early B-type stars in Fig. \ref{lxlbol}.
L$_{\rm x}$ values were calculated from our X-ray spectral analysis. 
Bolometric luminosities with relative uncertainties were estimated form the
known spectral types and the calibrations given by \citet{2005A&A...436.1049M}.

The proportionality between L$_{\rm x}$ and L$_{\rm bol}$ is retrieved, albeit
with a large dispersion. However, when we distinguish between main sequence
stars and evolved ones (i.e. giants and super-giants), we discover, for these
latter, an apparently tight L$_{\rm x}$ - L$_{\rm bol}$ relationship which is
not a simple proportionality. In fact, for evolved stars, the $\chi^2_\nu$
with respect to the L$_{\rm X}$=10$^{-7}$L$_{\rm bol}$ relation is 2.61, while
if we adopt the best-fit power law (with index $\Gamma$$\sim$1.8) the
$\chi^2_\nu$ is reduced to $\sim$1.3\,. We test the statistical significance
of  including this additional degree of freedom in the L$_{\rm x}$/L$_{\rm
bol}$  relation. We used a F$_\chi$-test as described in
\cite[e.g.][]{1969drea.book.....B}. We found that F$_\chi$-test $\sim$ 0.6,
which does not indicate a significant improvement of the quality of the fit
(F$_\chi$ should be greater than 1.1), even when adopting a significance level
as large as 0.1. However, this result is affected by the small number of
evolved O and B stars in our sample. A similar analysis with more stars from
other regions will be needed in the future to confirm, or disprove, this
tentative result.

A more detailed study of the physical origins of the X-ray emission from OB
stars and on its dependence on stellar wind parameters will be presented in
forthcoming paper.

\subsubsection{Specific sources}

We now discuss in some detail the results of our analysis for three  noteworthy
OB stars in the ACIS FOV that have been the subject of previous investigations.

\begin{figure*}[!ht] \centering
\includegraphics[width=3.75cm,angle=270]{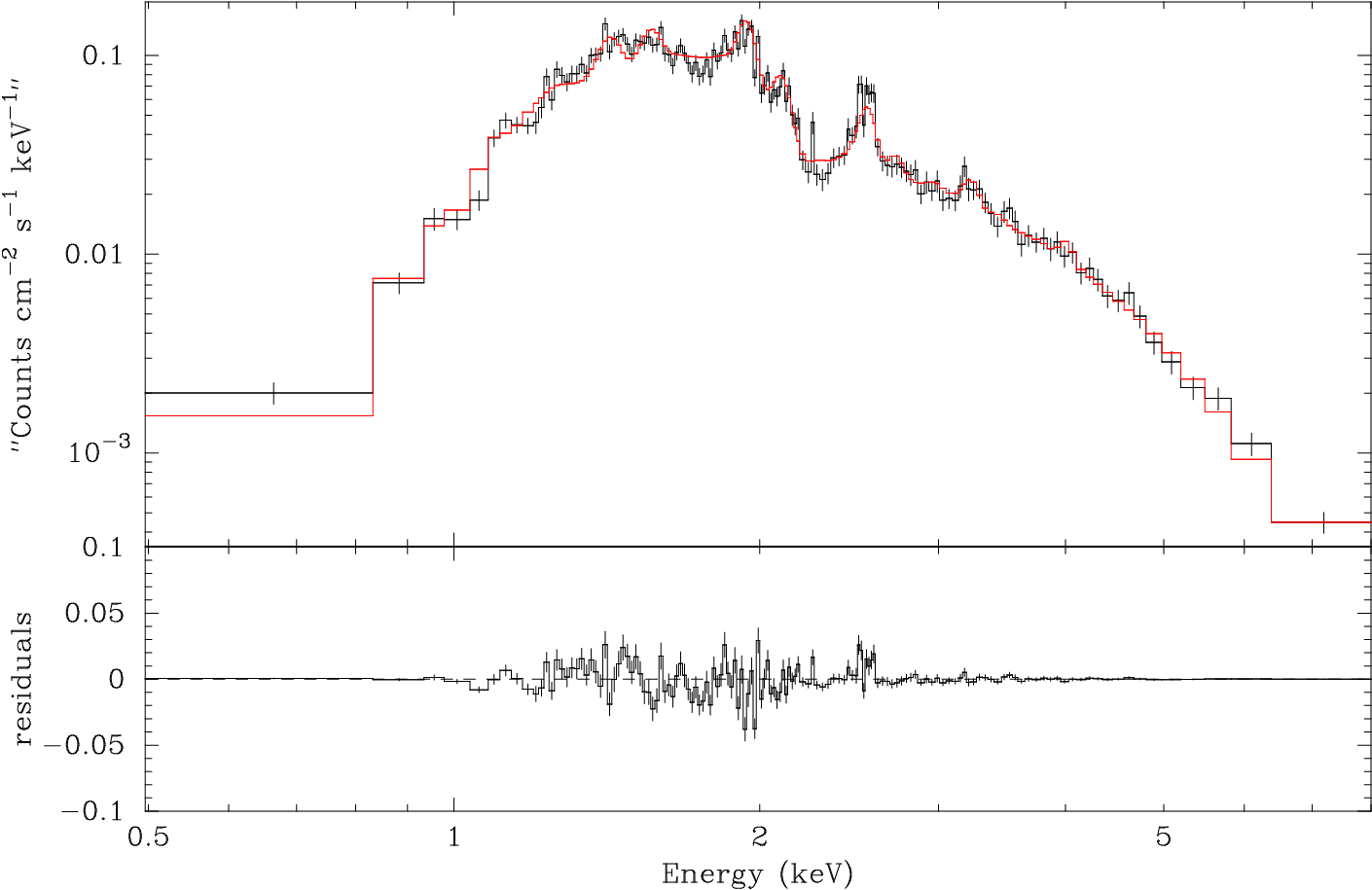}
\includegraphics[width=3.75cm,angle=270]{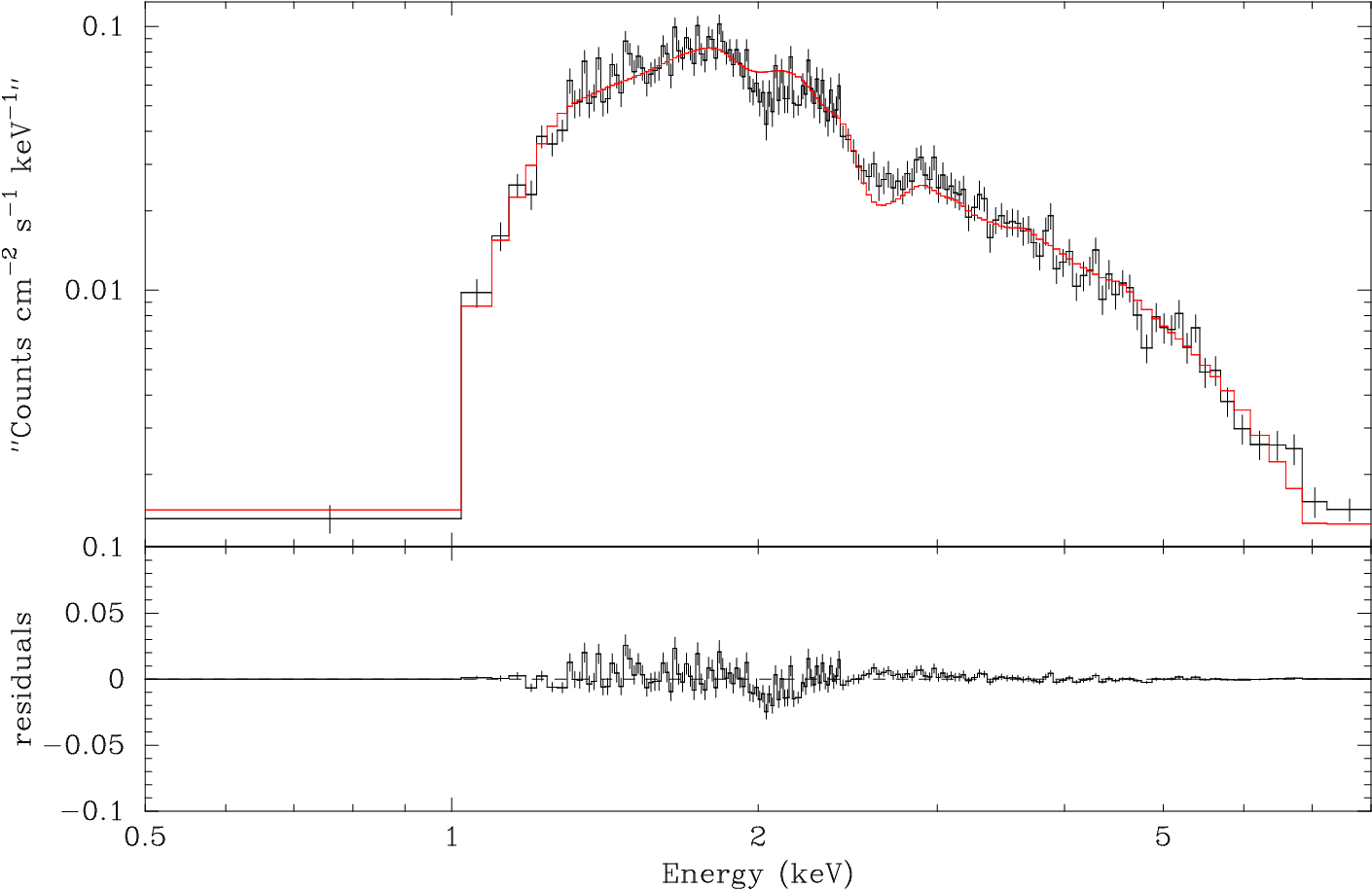}
\includegraphics[width=3.75cm,angle=270]{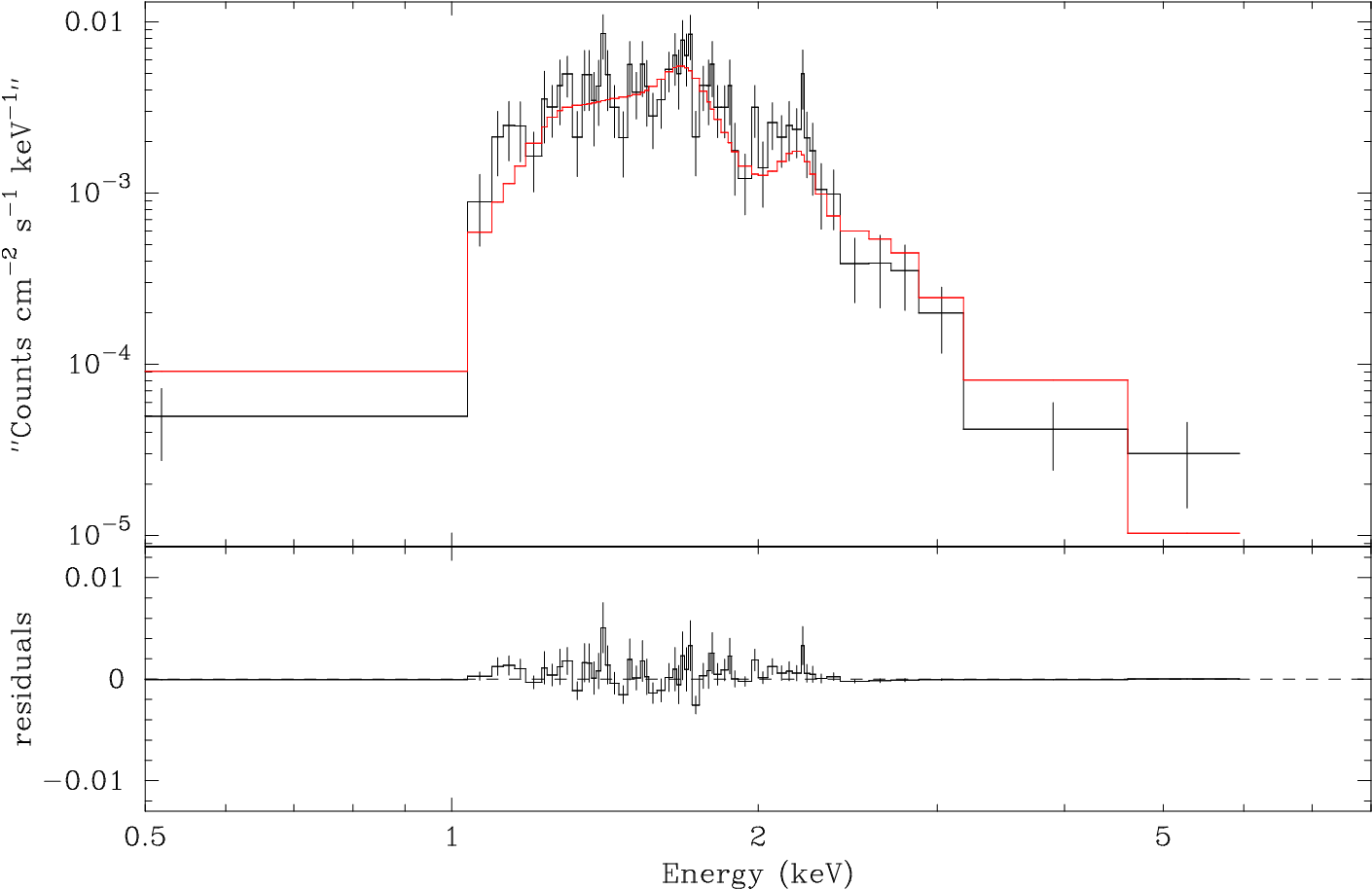} \caption{ACIS-I spectra
of three OB stars. The data were fitted with  absorbed two temperatures
optically thin thermal plasma models. Left: Cyg\#12 (Src.Id. \#60), showing a
prominent FeK 6.7 keV emission line, indicative of hot thermal plasma. Center:
Cyg\#8 (Src.Id. \#568), a colliding wind binary with a strong FeK$_\alpha$
blend at 6.4+6.7 keV,  arguing in favor of the thermal nature of the hard part
of the spectrum  and consistent with the expected emission from a colliding
wind region \citep{2003AdSpR..32.1161R}. Right: source \#625 (O5.5\,If), a
typical super-giant fit with a 1-T model with a soft temperature model:
KT$_1$=0.55$\pm$0.06 keV.} \label{obspectra} \end{figure*}

\begin{itemize}

\item {\bf \cyg\#8} (ACIS source \# 568). It has recently been classified as an
O6If\,+\,O5.5III(f) binary with a 21.9d period a strong evidence of
phase-locked X-ray variability \citep{2005mshe.work...73D}. Our observation
only covers $\sim$5\% of the orbital period and we do not detect variability.
\cyg\#8 has also been observed by \cite{2004ApJ...616..542W} with the {\em
Chandra} High Energy transmission Grating Spectrometer (HETGS). They fit the
X-ray spectrum with a a 2T thermal model with T$_1$=3.98$\pm$1.67 MK
(0.34$\pm$0.14 keV) and T$_2$ 13.88$\pm$1.80 MK (1.2$\pm$0.15 keV) and obtain
an L$_{\rm x}$/L$_{\rm bol}$ ratio of $\sim$1.8$10^{-7}$. The spectral
parameters are in rough agreement with those obtained here (see
Figure\,\ref{obspectra} and results presented in Table\,\ref{OB}),  but we find
a somewhat higher L$_{\rm x}$/L$_{\rm bol}$, $\sim$4.0$\times$$10^{-7}$. The
discrepancy is however due exclusively to the different adopted bolometric
luminosities: \cite{2004ApJ...616..542W} use log L$_{\rm bol}$\,=\,39.79 from
\cite{1989ApJ...340..518B}, while we here adopt log(L$_{\rm bol}$)\,=\,39.4
obtained from \cite{2002A&A...396..949H}.

\item {\bf \cyg\#9} (ACIS source \#\,488). An O5\,If super giant, it is the
strongest and most variable non-thermal radio emitter in \cyg 
\citep{2005PhDT.........1V}. In X-rays it is however fainter than other evolved
stars. \cyg\#9 has not been indicated as a binary in the literature. However,
periodic radio variability has been recently observed with a period of
$\sim\,$2.35 yr (Blomme, private communication).  \citet{1998ApJS..118..217W}
did not detect variability in {\em EINSTEIN} and {\em ROSAT} X-ray data.
\citet{2005mshe.work..103R} successfully fit the {XMM-Newton} EPIC X-ray
spectrum with an absorbed 2-T thermal model (kT$_1$\,=\,0.63$\pm$0.03 and
kT$_2$\,$\sim$\,2.4 keV). These results are roughly consistent with ours (c.f.
Table\,\ref{OB}).

\item {\bf \cyg\#12} (ACIS source \# 60), has been variously classified and its
spectral type is probably variable. Most investigations however agree with an
extremely luminous B supergiants B5\,Ie  \citep{2003ApJ...597..957H}, possibly
related to the  Luminous Blue Variable (LBV) phenomenon. Recent two-dimensional
spectral classification suggests a B5+/-0.5\,Ia$^+$ spectral type 
\citep{2004ARep...48.1005K}. In our near-IR color-color diagram (Fig.
\ref{jh_hk}), \cyg\#12 appears to be highly absorbed and to be affected by a
near-IR excess typical of Herbig AeBe stars \citep{1992ApJ...393..278L}.
\cite{2004ARep...48.1005K} present evidence of a line radial velocity gradient
that can be interpreted as an indication of matter infall. These findings
suggest the existence of an accretion disk around the star.

The {\em XMM-Newton} EPIC  spectrum of \cyg\#12  was also fitted by
\cite{2005mshe.work..103R},  who find temperatures kT$_1$\,=\,0.73$\pm$0.16 and
kT$_2$\,=\,1.8$\pm$0.4 keV, consistent with our own estimation (see
Figure\,\ref{obspectra} and Table\,\ref{OB}). The temperature of the hard
component is significantly higher than expected according to the wind shock
emission model, given the low velocity of the stellar wind, $\sim$\,150
km\,s$^{-1}$. Emission from MCWS might better explain the spectrum (see above).
Such a scenario could also explain the observed X-ray variability  (see \S
\ref{sect:var}) if the magnetosphere is tilted with respect to the rotation
axis as in the case of $\theta^1$Orionis\,C, \cite{2005ApJ...634..712G}.

\end{itemize}

\subsection{Intermediate-mass stars}

Intermediate-mass stars ($\rm 2.8<M/M_\odot<10$) are not expected to emit
X-rays because, unlike O- and early B stars, they do not drive strong stellar
winds nor possess outer convective zones that can sustain a dynamo mechanism
such as the one that is ultimately held responsible for X-ray activity in
low-mass stars. Several studies have however observed X-ray emission apparently
associated with late B- and A-type stars. Although its origin has been often
attributed to the coronal activity of unresolved late-type companions
\citep[e.g. ][]{2006A&A...452.1001S,2006astro.ph..5590S}, the matter is
controversial. The \cyg region has a large number of intermediate mass stars.
It is therefore an ideal target to check this hypothesis. We classify 66 ACIS
sources as intermediate-mass. The L$_{\rm X}$ vs. mass plot in Fig.
\ref{lxlbol} does not reveal any dramatic discontinuity at the boundaries with
low and high mass stars. Intermediate mass stars, however, do not seem to
follow the trend of increasing L$_{\rm X}$ with mass of lower mass stars, and
seem to have L$_{\rm X}$ values that span the whole range spun by these latter.
This observation is consistent with the companion scenario.

As further supporting evidence for this conclusion we note that 8 of the X-ray
sources associated with intermediate mass stars (about 12\% of the sample) are
variable, which is almost the same fraction found for low-mass stars (see
\S\,6.3). Moreover, their X-ray spectra are well represented by isothermal
models with median kT=2.6\,keV, also in agreement with values found for low
mass stars (see \S \ref{sect:lm_stars})

\begin{figure}[!h] \centering
\includegraphics[width=8.75cm,angle=0]{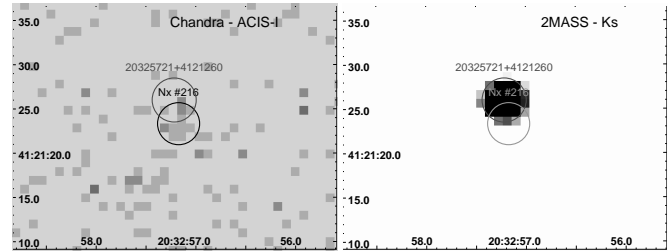} \caption{Example of a
possible misidentification of an X-ray source with a intermediate-mass near-IR
star. We show the X-ray and K$_{\rm s}$-band images of ACIS source \#216. The
X-ray and 2MASS positions differ by  $\sim$2.5 arcsec
($\sim$1.8$\times10^{-2}$\,pc at the \cyg distance). The X-ray source might be
instead associated with a faint near-IR source undetected in 2MASS.}

\label{AB}
\end{figure}

At the distance of \cyg an offsets between IR and X-ray positions of 1", i.e.
roughly the {\em on-axis} \chandra spatial resolution, corresponds to a
projected separation of 1450 AU. X-ray source positions can however be
determined with a precision of $\sim$0.1" (145 AU) or even better, depending on
the source statistics and off-axis. If the intermediate-mass primary of a
binary system is X-ray dark we should thus be able to distinguish, at least
statistically, whether the observed X-ray emission originates from the low mass
secondary in the case of {\em wide} binary systems, defined as systems with
separation $>$ 0.001pc (206 AU) \citep[cf.][]{1990AJ....100.1968C}.
Figure\,\ref{AB} shows an example of an X-ray source associated with an
intermediate-mass counterpart, but with a significant spatial offset. A
comparison of the distributions of X-ray-NIR offsets for X-ray sources
associated with low and intermediate mass counterparts is however inconclusive.
We conclude that, if the companion hypothesis is correct, most X-ray emitting
low-mass companions are not in wide systems. 

\subsection{Low-mass stars}
\label{sect:lm_stars}

The origin of X-ray activity in low-mass PMS stars has so far eluded full
understanding. Although many aspects of MS coronal activity are also not
understood, the X-ray activity of MS stars correlates well with stellar
rotational and convection properties, thus supporting the picture of a 
solar-like corona that is ultimately powered by a stellar dynamo. For 
Pre-Main Sequence (PMS) stars, however, no such a correlation has been found
and the picture appears somewhat complicated by the presence of accretion
disks,  which seem to play an important role in the X-ray emission. On one hand
plasma heated by shocks due to the accretion process is likely to be a positive
source of soft X-rays \citep[cf.][]{2002ApJ...567..434K,
2004A&A...418..687S,2005ApJS..160..469F,2006astro.ph..4243F}. On the other hand
the presence of accretion disks seems to lower the overall activity level,
although probably making the X-ray emission harder and more time variable
\citep[][]{2003ApJ...582..382F,2005ApJS..160..401P,2006astro.ph..4243F}.

We have defined the low mass range  as M$<$2.8M$_\odot$, based on the presence
of a significant convection envelope. The X-ray spectra and luminosities of low
mass stars in \cyg are quite typical of T-Tauri stars in other regions: our
spectral analysis (\S \ref{sect:specanal}) yields a median kT=2.38 keV and
log\,N$_{\rm H}$=22.23. Figure \ref{lx-mass} indicates that, although the
spread of L$_{\rm x}$ at any given mass is of the order of 1~dex,  a trend of
increasing L$_{\rm x}$ with mass is observed. The same figure shows for
comparison the nearly complete sample of Orion Nebular Cluster (ONC) stars as
observed by the Chandra Orion Ultra-deep Project (COUP)
\citep{2005ApJS..160..319G}. The sensitivity limits of the two observations are
quite different because of the larger distance of \cyg ($\sim$1450 pc vs. 470
pc for the ONC), and the longer exposure time of the COUP observation
($\sim$838  vs. $\sim$97 ksec for our data).  However, as already discussed,
the most severe limit to the completeness of our sample of likely members with
near-IR counterparts is likely to be the  sensitivity limit of the 2MASS data.
Note indeed that the \cyg stars in Fig. \ref{lx-mass} have a well defined limit
in mass, at $\sim0.4-0.5$M$_\odot$, rather than in X-ray luminosity.

Figure \ref{lx-mass} shows that the luminosities of low mass stars in \cyg and
in the ONC are fully compatible. We note here that the numbers of stars in the
1.0-2.8$M_\odot$ mass range in the \cyg and ONC regions differ considerably. In
\cyg we count 334 such stars, about 10 times as many as the in a sample of 375
COUP-detected lightly absorbed (\Av$<$2) ONC stars with mass estimates, in
which we find 34 stars in the same range \citep{2005ApJS..160..319G}.  Now,
assuming that the IMFs in the two regions are the same and that the considered
sample of lightly absorbed ONC stars is complete, we can estimate the total
number of stars in our \cyg ACIS FOV as 375/34$\times$334$\sim$3700.
Considering that our FOV contains only one fourth of the 106 OB stars known in
Cygnus OB2, we can then scale this number by the number of OB stars to get the
total expected \cyg population: $\sim$15000 stars. The total stellar mass of
the cluster can be then estimated by multiplying this number by the average
stellar mass in the lightly absorbed COUP sample (0.84\,M$_\odot$): M$_{\rm
cl}\sim 1.2\times 10^4$M$_\odot$, somewhat lower than the estimate of
\citet{2000A&A...360..539K}, who derived 4-10 $\times 10^4$M$_\odot$.

Deeper and more extended X-ray and near-IR surveys of \cyg will be invaluable
for the study of young stars of all masses with un-precedented statistic.

\section{Summary and conclusions}

We have reported the results of a deep {\em Chandra} X-ray observation pointed
toward the $\sim$2 Myr old star forming region Cygnus OB2. Source detection was
performed using PWDdetect, a wavelet-based algorithm, supplemented by visual
inspection, identifying 1003 X-ray sources in the 17'$\times$17' ACIS-I field
of view. Data extraction was performed using the semi-automated IDL-based {\sc
ACIS Extract} package, which is well suited to the analysis of observations of
crowded fields such as ours.

The X-ray source list was cross-identified with optical and near-IR (2MASS)
catalogs: 26 X-ray sources were identified with optically characterized OB
members of \cyg (out of the 33 lying in the FOV) and 775 with 2MASS sources.
Among these latter sources almost all are believed to be \cyg members. About
$\sim$80 X-ray sources without optical/NIR counterpart are estimated to be of
extragalactic nature (AGNs) while the remaining X-ray sources with no
counterpart are likely associated with members that are fainter than the 2MASS
completeness limit.

In order to characterize the previously unidentified likely cluster members
with NIR counterpart, we placed them on NIR color-magnitude (K$_{\rm
s}$\,vs.\,H-K$_{\rm s}$) and color-color  (H-K$_{\rm s}$\,vs.\,J-H) diagrams. A
first estimate of interstellar extinction was obtained adopting a 2\,Myr
isochrone for the  low- and intermediate-mass stars and assuming that  O- and
early B-type stars lie on the MS. We find a median visual absorption for OB
stars of \Av$\sim$5.6 mag, while low mass likely members seem to be slightly
more absorbed, \Av$\sim$7.0 mag. We also use the 2\,Myr isochrone and the J
magnitude to estimate masses of likely members assuming that they share the
same distance and absorption. Our sample of X-ray selected members with near-IR
counterparts reaches down to M=0.4-0.5\,M$_\odot$, and is likely complete down
to $\sim$1\,M$_\odot$. From the H-K$_{\rm s}$\,vs.\,J-H diagram we estimate the
fraction of low-mass stars with NIR excesses (23/519), finding it to be quite
low with respect to the 1 Myr old ONC region indicating that either (i) the
disk fraction decreases steeply with stellar age, or (ii) the intense
photo-evaporating UV-field due to the large population of O stars in the region
has reduced the disk lifetime. 

At least 85 or 134 X-ray sources, i.e. $\sim$8.5\% and 13\% of the total,  were
found to be variable within our observation with a confidence level of 99.9\%
or 99\%, respectively.  The fraction of variable sources increases with
increasing source counts, likely as a results of a statistic-induced bias. The
light-curves of 97 sources indicate an impulsive, flare-like, behavior while
the rest show more gradual variations of the X-ray emission. Only two of the 24
detected O- and early B-type stars are detected as variable during our 97 ksec
observation, in spite of the high statistics of their light-curves. These
exceptions are the B5\,Ie star Cygnus\#12 (our source \#60), showing a rather
linear decay of the count rate during the observation, and a B1.5\,V+$?$ star
(\#979), which also shows a slow variability, maybe related to rotational
modulation. In this latter case the detected emission may originate from the
primary B1.5 component and/or from its presumably lower mass companion
indicated by the spectral type published by  \citet{2003ApJ...597..957H}.

We modeled the ACIS X-ray spectra of sources with more than 20 photons
assuming absorbed thermal emission models. Median log(N$_{\rm H}$) and kT
values of the sources are 22.25 (cm$^{-2}$) and 2.36 keV, respectively. 
Variable sources, P$\rm _{KS}<0.01$ have harder spectra (median kT: 3.3 keV)
respect to non-variable ones (median kT:2.1 keV) and flaring sources are even
harder (median kT:3.8). Sources associated with O- and early B-type stars are
instead quite soft (median kT: 0.75 keV). Absorption corrected X-ray
luminosities of OB stars were calculated from the best fit spectral models. For
the other, typically fainter, sources, we adopted a single count-rate to flux
conversion factor computed from the results of individual spectral fits.  Low
mass stars have L$_{\rm x}$'s ranging between $10^{30}$ and $10^{31}$
erg\,s$^{-1}$ (median $2.5\times10^{30}$). Variable low mass stars are on
average 0.4 dex brighter, probably because of a selection effect. These X-ray
luminosities are consistent with those of similar mass stars in the slightly
younger (1\,Myr) ONC region. O and B-type stars are the most luminous, with
L$_{\rm x}$= 2.5$\times10^{30}$-6.3$\times10^{33}$  erg\,s$^{-1}$. Their X-ray
and bolometric luminosities are in rough agreement with the relation  L$_{\rm
x}$/L$_{\rm bol}=10^{-7}$, albeit with an order of magnitude dispersion. The
X-ray luminosities of the 10 {\em evolved} OB stars in our sample however, seem
to obey a better defined and steeper power-law relation with $\rm L_{bol}$: 
L$_{\rm x}$$\propto$L$_{\rm bol}^{1.8}$). Further studies of a larger number of
evolved OB-type stars are needed to confirm of disprove this dependence.

\begin{acknowledgements}

This publication makes use of data products from the Two Micron All Sky Survey,
which is a joint  project of the University of Massachusetts and the Infrared
Processing and Analysis  Center/California Institute of Technology, funded by
the National Aeronautics and Space  Administration and the National Science
Foundation. J.F.A.C acknowledges support by the Marie Curie Fellowship Contract
No. MTKD-CT-2004-002769 of the project "{\it The Influence of Stellar High
Energy Radiation on Planetary Atmospheres}", and the host institution INAF -
Osservatorio Astronomico di Palermo (OAPA). E.\,F., G.\,M., F.\,D. and S.\,S.
acknowledge financial support from the Ministero dell'Universita' e della
Ricerca research grants, and ASI/INAF Contract I/023/05/0.

\end{acknowledgements}

\bibliographystyle{astron}
\bibliography{Albacete_Colombo_rev}

\end{document}